\documentclass{jfm}
\usepackage{caption}
\usepackage{subcaption}
\usepackage{graphicx}
\usepackage{epstopdf, epsfig}
\usepackage{bbold}
\usepackage{siunitx}
\usepackage{cmll}
\usepackage[mathcal]{eucal}
\usepackage{ucs}
\usepackage[utf8x]{inputenc}
\usepackage{amsmath}
\def\bm#1{\mbox{\boldmath{$#1$}}}
\def\rr#1{(\ref{#1})}

\newcommand{\be}{\begin{equation}}
\newcommand{\ee}{\end{equation}}

\shorttitle{Evanescent waves}
\shortauthor{E. Kirkinis and M. Olvera de la Cruz
}
\title{Evanescent and inertial-like waves in rigidly-rotating odd viscous liquids}
\author{E. Kirkinis
\corresp{\email{kirkinis@northwestern.edu} }
\and M. Olvera de la Cruz }
\affiliation{Department of Materials Science \& Engineering, Robert R. McCormick School of Engineering and Applied Science, Northwestern University , Evanston IL 60208 USA\\ Center for Computation and Theory of Soft Materials, Northwestern University , Evanston IL 60208 USA
}

\begin{document}

\maketitle

\begin{abstract}
Three-dimensional non-rotating odd viscous liquids
give rise to Taylor columns and support {axisymmetric} inertial-like waves [\emph{J. Fluid Mech.}, vol. {973}, A30, (2023)]. When an odd viscous liquid is subjected to 
rigid-body rotation however, there arise in addition a plethora of other phenomena that need to be clarified. In this paper
we show that three-dimensional incompressible or two-dimensional compressible odd viscous liquids, rotating rigidly with angular velocity $\Omega$, give rise to both oscillatory and evanescent inertial-like waves or a combination thereof (which we call of mixed type), that can be \emph{non-axisymmetric}. 
By evanescent we mean that along
the radial direction, typically when moving away from a solid boundary, the velocity field decreases exponentially. These waves precess in a prograde or retrograde manner with respect to the rotating frame. The oscillatory and evanescent waves resemble, respectively, the body and wall-modes observed in (non-odd) rotating Rayleigh-B\'enard convection [\emph{J. Fluid Mech.}, vol. {248}, pp. 583-604 (1993)]. We show that the three types of waves 
(wall, body or mixed) can be classified with respect to pairs of
planar wavenumbers $\kappa$ which are complex, real or a combination, respectively. Experimentally, by observing the precession rate of the patterns, it would be possible to determine the largely unknown values of the odd viscosity coefficients. 
This formulation recovers as special cases recent studies of equatorial or topological waves in two-dimensional odd viscous liquids which provided examples of the bulk-interface correspondence at frequencies $\omega<2\Omega$. 
We finally point out that the two and three-dimensional problems are 
formally equivalent. Their difference then lies in the way data propagate along characteristic rays in three dimensions, 
which we demonstrate by
classifying the resulting Poincar\'e-Cartan equations. 
 \end{abstract}

\section{Introduction}
Odd viscous liquids are dissipationless in the sense that they do not give rise to viscous heating \citep{Landau1987}. 
They were systematically studied by \citet{Avron1998}, following previous work on the quantum Hall effect
\citep{Avron1995}. Their constitutive laws however, were already known in the context of polyatomic gases
where a detailed experimental and theoretical program was carried-out at Leiden with a \emph{terminus ante quem} in the 1960's
\citep{Beenakker1970,Hulsman1970}.
Recent experiments established the existence of odd viscosity in active liquids,
which acted to suppress surface undulations in a manner resembling surface tension
\citep{Soni2019}. In addition to the odd viscosity coefficients cited in the above works, there are others 
that may appear in materials endowed with discrete symmetries cf. \citep{Rao2020,Souslov2020}. The review article by \citet{Fruchart2023} discusses the above experiments and various physical 
effects that arise in the presence of odd viscosity. 
%

A previously observed odd viscosity-induced uncommon physical effect, related to the corpus of the 
present paper is the propagation of inertial-like waves in a three-dimensional odd viscous liquid.  
This is the case because such a liquid is endowed with an intrinsic
mechanism that tends to restore a fluid particle back to its equilibrium position. 
In addition, a body moving slowly in a quiescent three-dimensional odd viscous liquid
will be accompanied by a liquid cylindrical column whose 
generators circumscribe the body \citep{kirkinis2023axial}. 

\begin{figure}
\vspace{5pt}
\begin{center}
\includegraphics[height=2.2in,width=3.9in]{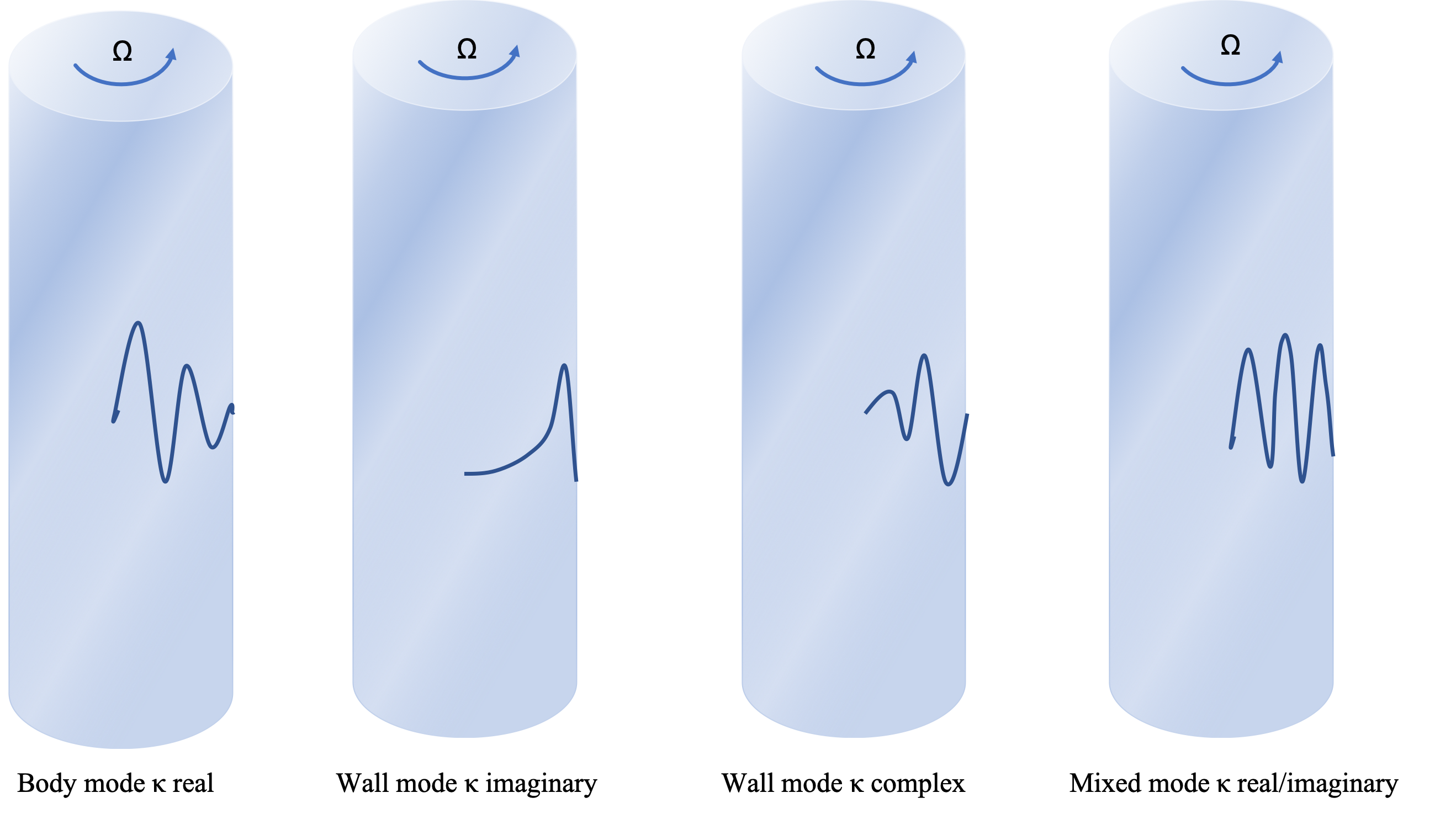}
\vspace{-0pt}
\end{center}
\caption{
\label{body_wall_cartoon}Wall and body modes (or evanescent and oscillatory inertial-like waves, respectively), and mixed mode of the fields (density or pressure $\sim J_m(\kappa r)$, where $J_m$ is the Bessel function of the first kind and $m$ an integer determining periodicity in the azimuthal direction), for a rigidly-rotating odd viscous liquid with angular velocity $\Omega$,  in two and three dimensions, satisfying no-slip boundary conditions. Each mode can be classified according to the character of the planar wavenumber $\kappa$, cf.  table \ref{tab: roots}. \emph{Body modes:} Prominent in the interior of the cylinder. 
\emph{Wall modes:} Prominent near the side wall. \emph{Mixed modes:} A combination of the previous two behaviours. }
\vspace{-0pt}
\end{figure}

The main result of this paper is the determination of wall and body-like modes in a rigidly-rotating odd viscous liquid with angular velocity $\Omega$ in a disk or a cylinder of radius $R$ with no-slip boundary conditions, that resemble 
their non-odd counterparts in rotating Rayleigh-B\'enard convection \citep{Goldstein1993,Knobloch1994}. We identify the wall
modes with evanescent waves and the body modes with oscillatory inertial-like waves. Both types can be classified with respect to a single 
planar wavenumber $\kappa$, which is complex or real, respectively cf. Fig. \ref{body_wall_cartoon}.
These waves precess in a prograde or retrograde manner with respect to the rotating frame. Wall modes are prominent close to a solid boundary and body modes in the interior of the cylinder (or disk). An odd viscous liquid provides a third case where admissible wavenumbers $\kappa$ are concurrently real and imaginary, that we call ``mixed'' in this paper. The classification of the physical behaviours according to the character of $\kappa$ is displayed in table \ref{tab: roots}.

Following the theory developed in \citep{Goldstein1993,Knobloch1994} we obtain the (exact) fields by satisfying the 
(no-slip in this paper) boundary conditions. This method provides the admissible curves in the parameter space spanned by the precession frequency $\omega$ and 
odd viscosity coefficient $\nu_o$, giving rise to the aforementioned behaviour. Thus, it is possible to theoretically determine the largely unknown values of the odd viscosity coefficients by 
experimentally observing the frequency $\omega$ of the precessing patterns.

Recent studies on equatorial  \citep{Tauber2019} and topological waves \citep{Souslov2019} in two-dimensional odd viscous liquids
provided examples of the bulk-interface correspondence by establishing the presence of topological waves at frequencies $\omega<2\Omega$.  Our formulation recovers these effects as special cases.

This paper is thus organised as follows. 
In section \ref{sec: inertialc} we formulate the  \emph{non-axisymmetric} motion of a two-dimensional compressible rigidly-rotating odd viscous liquid in a disk geometry of radius $R$ following the arguments developed earlier for (non-odd) rotating Rayleigh-B\'enard convection \citep{Goldstein1993,Knobloch1994} and in \citep{Chandrasekhar1961}. The variable part of the density $\rho'$, satisfies a scalar Poincar\'e-Cartan equation which leads to a relation between the 
planar wavenumber $\kappa$, material parameters and precession frequency $\omega$ in parameter space. 
Only one of these wavenumbers $\kappa$ is however admissible; it can be determined by solving a secular equation 
obtained by satisfying the (no-slip here) boundary conditions on the side-wall. Thus, the density profiles so obtained, 
are precessing with frequency $\omega$ in the rotating frame of the disk and can be exponential or oscillatory in the radial direction. The radially exponential
profiles are evanescent waves and resemble the wall modes obtained in (non-odd) rotating Rayleigh-B\'enard convection \citep{Goldstein1993,Knobloch1994}, although the wall modes of the latter system are a consequence of thermal
forcing and supercritical behaviour of a non-odd system endowed with shear viscosity. The radially exponential
profiles can also be understood as equatorial  \citep{Tauber2019} and topological waves \citep{Souslov2019}
(that is, waves that propagate parallel to a boundary and decay away from it exponentially). 
Solution of the real and imaginary parts of the secular equation gives parameteric curves of admissible 
$(\omega, \nu_o)$ values (where $\nu_o$ is the coefficient of kinematic odd viscosity) leading to the aforementioned
exponential/wall mode/evanescent behaviour. Therefore, this formulation can provide the means of determining the 
largely unknown odd viscosity coefficient by observing the precessing rate of patterns in an experiment.  
The special case of the axisymmetric $m=0$ mode is relegated to the Supplementary Materials addendum that includes a number
of illustrative examples associated with this mode. 

%
%
%

In section \ref{sec: inertial} we formulate the non-axisymmetric motion of a  \emph{three-dimensional incompressible} rigidly-rotating odd viscous liquid in a cylinder of radius $R$. The formulation is nearly identical to the two-dimensional case of section \ref{sec: inertialc} with the exception of the presence of an axial velocity component $v_z(r,\phi,z,t)$, whose arguments are expressed with respect to cylindrical coordinates, and two (rather than one) odd viscosity coefficients $\nu_o$ and $\nu_4$. A Poincar\'e-Cartan and a secular equation give rise to the admissible planar real or complex wave-numbers 
$\kappa$ leading to precessing body and wall modes, respectively. Since the patterns precess in the rotating frame, again, one could determine the unknown odd viscosity coefficients by experimentally observing their rotation rate. 
The observation of evanescent waves in rigidly-rotating (non-odd) inviscid or viscous liquids is rather rare with the 
exception of a recent experimental study \citep{Nosan2021}, and references therein.
We thus adapt the experimental conditions of \citet{Nosan2021} to the case of a three-dimensional odd viscous liquid
rotating rigidly with angular velocity $\Omega$. We show that fluid particle paths are ellipses lying on $r-z$ planes
and can possibly be employed to determine the unknown values of the odd viscosity coefficients.

The two and three-dimensional problems are formally equivalent and the density $\rho'$ of the former plays the same role as the axial velocity 
component $v_z$ in the latter, as discussed in section \ref{sec: equivalence}. 
This behaviour is related to the conservation of helicity which is a consequence of the alignment of velocity with vorticity.
This tendency of the two fields to alignment, even when \emph{nonlinear} terms of the Navier-Stokes equations
are included, is expected from general grounds \citep{Pelz1985}. 
We relegate this discussion to the Supplementary Materials addendum.

The effects described in the main body of this paper are affected by the lower order terms of the governing partial differential
equations (the dispersion relation). We show in Appendix \ref{sec: Poincare} how higher order terms are responsible for the 
propagation of data in directions explicitly determined by the values taken by the odd viscosity coefficients.

\begin{figure}
\vspace{-5pt}
\begin{center}
\includegraphics[height=2in,width=2.5in]{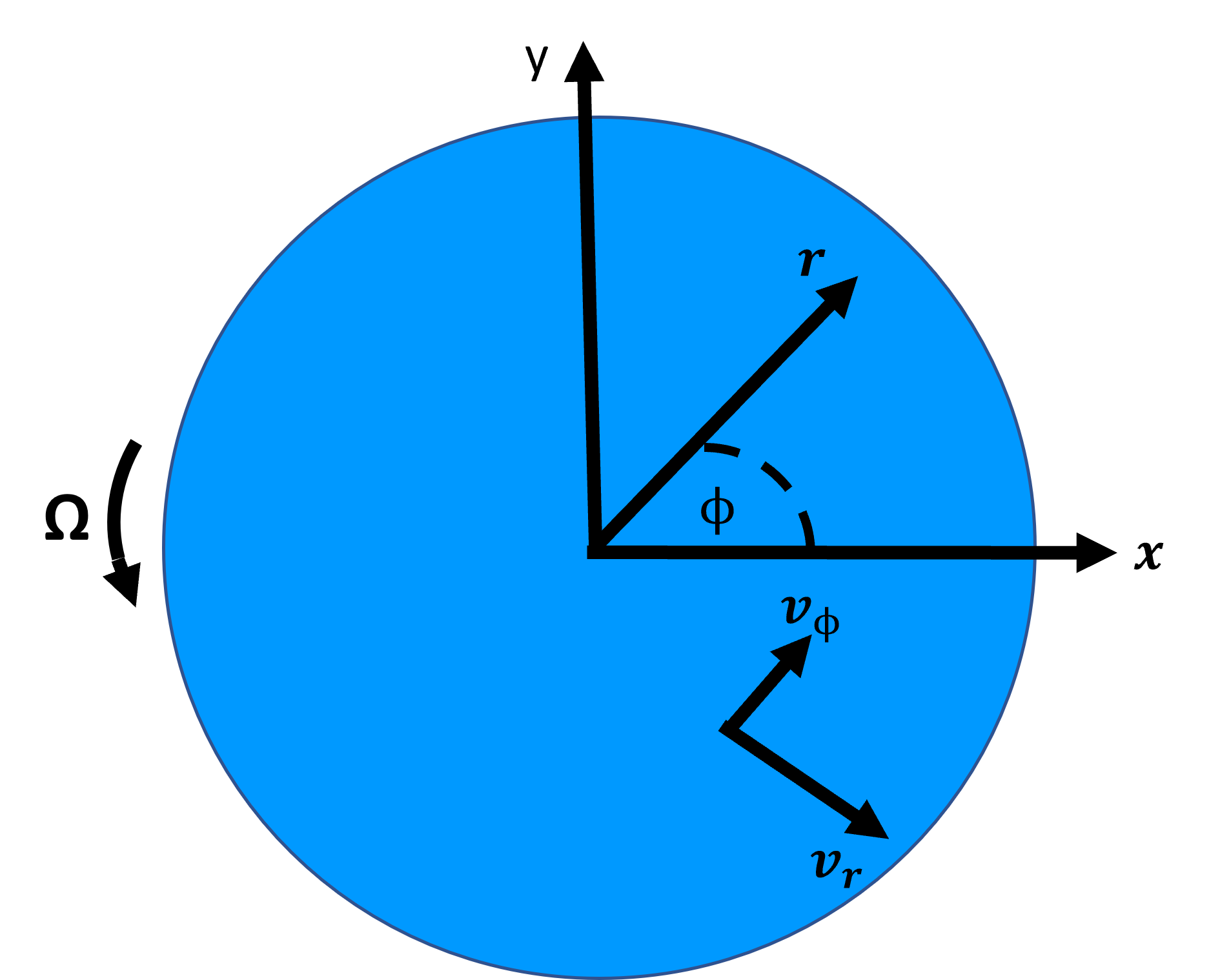}
\vspace{-0pt}
\end{center}
\caption{Two-dimensional odd viscous compressible liquid rotating with angular velocity $\Omega$. In plane polar coordinates the velocity field is 
$\mathbf{v} = v_r \hat{\mathbf{r}} + v_\phi \hat{\bm{\phi}}$ in the 
frame rotating with the liquid at constant angular velocity $\Omega$. 
\label{disk}  }
\vspace{-0pt}
\end{figure}

\section{Evanescent and inertial-like oscillations in a rigidly-rotating compressible two-dimensional odd viscous liquid}
\subsection{\label{sec: inertialc}Non-axisymmetric waves}
A two-dimensional odd viscous liquid obeys the constitutive law 
\citep{Lapa2014, Ganeshan2017,Banerjee2017} 
\be \label{sigma0c}
\bm{\sigma}' = \eta_o 
\left(\begin{array}{cc}
-\left(\partial_r v_\phi - \frac{1}{r}v_\phi + \frac{1}{r}\partial_\phi v_r \right) 
&  \partial_r v_r - \frac{1}{r}v_r - \frac{1}{r}\partial_\phi v_\phi \\
\partial_r v_r - \frac{1}{r}v_r - \frac{1}{r}\partial_\phi v_\phi  &  \partial_r v_\phi - \frac{1}{r}v_\phi + \frac{1}{r}\partial_\phi v_r  
\end{array}
\right)
\ee
where $\eta_o$ is termed the odd viscosity coefficient. We here consider a two-dimensional compressible
liquid rigidly-rotating with angular velocity $\Omega$, endowed with the above constitutive relation and satisfying
the continuity equation
\be
\partial_t \rho' + \rho\textrm{div} \mathbf{v} =0 \quad \textrm{for} \quad \rho'\ll\rho,
\ee
where $\rho'$ is the variable part of the density and $\rho$ a constant background level. 
In plane polar coordinates $r, \phi$ (cf. Fig. \ref{disk}) consider azimuthally-dependent fields of the form 
\be \label{vrhophit}
\left[v_r(r), v_\phi(r), \rho'(r)\right] e^{i(m\phi  -\omega t)}
\ee
where $\omega$
is a real frequency and $m$ an integer. Employing the 
constitutive law \rr{sigma0c}, the linearized equations of motion and continuity, in the frame of reference rotating with the liquid \citep[\S89]{Landau1981}
become
\begin{align} \label{nsrc}
-i\omega v_r & = -\frac{c^2}{\rho}\frac{\partial \rho'}{\partial r} + 2\Omega v_\phi 
- \nu_o\left[ \mathcal{L} v_\phi + \frac{2im}{r^2} v_r\right],  \\
-i\omega v_\phi & = -\frac{im c^2}{r}\frac{\rho'}{\rho}-2\Omega v_r +
\nu_o\left[ \mathcal{L} v_r - \frac{2im}{r^2} v_\phi \right],  
\label{nsphic}
\\
-i\omega \rho'& = -\rho \left[ \frac{1}{r} \frac{\partial}{\partial r} \left( r v_r\right) + \frac{im}{r} v_\phi \right], \label{nszc}
\end{align}
where $c$ is the speed of sound, $\nu_o=\eta_o/\rho$ and $\mathcal{L}$ is the linear operator
\be \label{L}
\mathcal{L} = \nabla_2^2 -\frac{1}{r^2},
\ee
$\nabla^2_2$ is the two-dimensional Laplacian $\frac{1}{r} \frac{\partial}{\partial r} \left( r \frac{\partial }{\partial r}\right)   - \frac{m^2}{r^2}$ and we neglected the nonlinear terms assuming small-amplitude motions. 

\begin{figure}
\vspace{-5pt}
\begin{center}
\includegraphics[height=2.6in,width=3.6in]{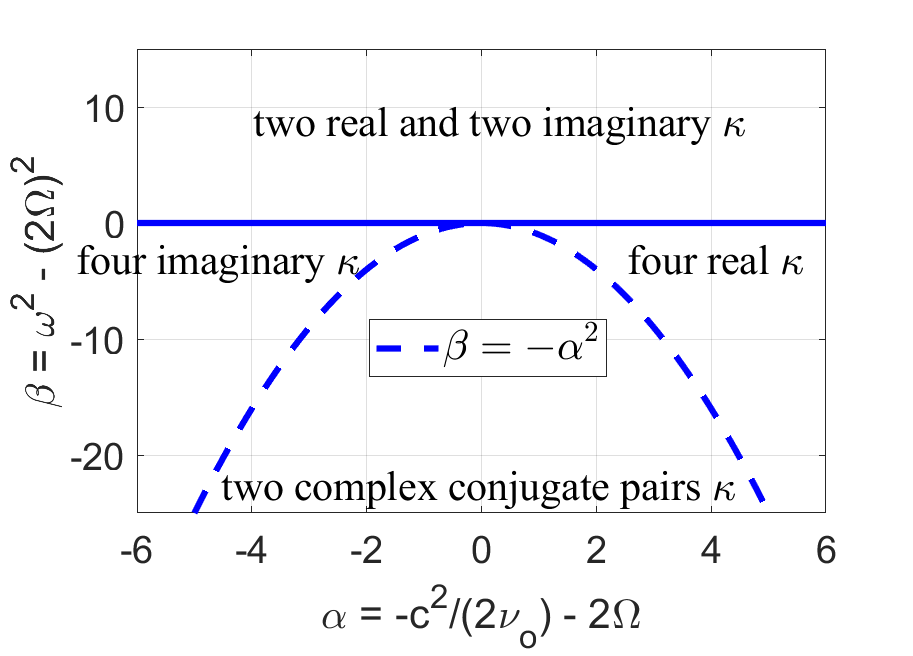}
\vspace{-0pt}
\end{center}
\caption{
\label{roots2D}Roots of Eq. \rr{kappa2c} in the parameter space $(\alpha,\beta)$ defined in Eq. \rr{alphabetac}
giving rise to the wall and body modes depicted in Fig. \ref{body_wall_cartoon} according to whether $\kappa$ 
is real, imaginary or complex. $\alpha$ and $\beta$ have units of frequency and square frequency, respectively.}
\vspace{-0pt}
\end{figure}

We impose no-slip boundary conditions at $r=R$
\be \label{bc2D}
v_r(r=R) = v_\phi(r=R) =0. 
\ee
These can be relaxed and replaced by mixed no-slip and force-free boundary conditions as reported in 
\citep{Souslov2019}. 

In Appendix \ref{sec: Poincarec} we reduce the momentum and continuity equations into a single equation for the density $ \rho'$
\be \label{PC2Db}
\partial_t\left[ \left(2\Omega - \nu_o\nabla_2^2\right)^2 - c^2 \nabla_2^2 + \partial_t^2   \right] \rho' =0.
\ee
Substituting 
\be
\rho'(r,\phi) = J_m(\kappa r) e^{i(m\phi -\omega t)}
\ee
into \rr{PC2Db}, where $J_m$ is the Bessel function of first kind, we obtain 
the relation
\be\label{disp2}
-\kappa^{4} \nu_o^{2}-4 \Omega  \,\kappa^{2} \nu_o -c^{2} \kappa^{2}-4 \Omega^{2}+\omega^{2} =0,
\ee
satisfied by the (possibly complex) wavenumber $\kappa =\kappa(c, \Omega,\omega, \nu_o)$, where all
parameters in the round brackets are real.  
Solving \rr{disp2} for $\kappa^2$ leads to 
\be \label{kappa2c}
\nu_o\kappa^2 = 
\alpha \pm \sqrt{\alpha^2 + \beta},
\ee
where 
\be \label{alphabetac}
\alpha = -\frac{c^2}{2\nu_o} - 2\Omega, \quad \beta = \omega^2 - (2\Omega)^2. 
\ee
Therefore $\kappa$ in Eq. \rr{kappa2c} can be real, imaginary or complex as displayed in, Fig. \ref{roots2D} and Table \ref{tab: roots}. Each one of these three root types thus corresponds to the three density or pressure behaviours
depicted in Fig. \ref{body_wall_cartoon}.

\begin{table}
  \begin{center}
\def~{\hphantom{0}}
  \begin{tabular}{|c|c|c|c|c|}
$\alpha^2 + \beta$& $\alpha$& $\beta$& Type of root $\kappa$&Physical effect\\
\hline
$+$&&$+$ & two real, two imaginary& Mixed modes\\
$+$&$+$&$-$ & four real& Body modes\\
$+$&$-$& $-$& four imaginary& Wall modes\\
$-$&& & two complex conjugate pairs& Wall modes\\
  \end{tabular}
\caption{Types of roots $\kappa$ displayed in Fig. \ref{roots2D}, from Eq. \rr{kappa2c} $\nu_o\kappa^2 = 
\alpha \pm \sqrt{\alpha^2 + \beta}$ according to the sign
of the parameters $\alpha = \frac{\omega^2}{2\Omega_o} - 2\Omega$ and $\beta= \omega^2 - (2\Omega)^2$ defined in Eq. \rr{alphabetac}. The last column defines the terminology employed in this paper to describe the physical effect}
  \label{tab: roots}
  \end{center}
\end{table}

From this point onwards we follow the solution method employed by  \citet{Goldstein1993} in determining the fields arising in non-odd rapidly-rotating Rayleigh-B\'enard convection. From the four values of $\kappa$ obtained in \rr{kappa2c}, only two give rise to linearly independent solutions. 
Let 
$\kappa_1$ and $\kappa_2$ denote these $\kappa$ values. The fields can then be stated as 
\be \label{2Dsol}
\left(
\begin{array}{c}
v_r \\ v_\phi
\end{array}
\right) = \sum_{j=1}^2
A_j\gamma_j \left(
\begin{array}{cc}
\delta_j  & 2\Omega \\
-2\Omega&  \delta_j
\end{array}
\right) \left(
\begin{array}{c}
\partial_r \\ \frac{im}{r}
\end{array}
\right)J_m(\kappa_j r), \quad \rho' = \sum_{j=1}^2A_jJ_m(\kappa_j r)
\ee
where, the coefficients $\gamma_j$ and $\delta_j$ are 
\be
\gamma_j = \frac{\kappa_j^{2} \nu_o +2 \Omega}{2 \Omega  \,\kappa_j^{2} \rho}, \quad 
\delta_j = \frac{-2 i \omega \Omega}{\nu_o \kappa_j^{2}+2 \Omega}, \quad j=1,2,
\ee
determined by substituting the solution \rr{2Dsol} into the continuity equation \rr{nszc} and into the $z$ component 
of the vorticity equation, and $A_j$ are complex constants determined by the boundary conditions. 
Substituting \rr{2Dsol} into the boundary conditions \rr{bc2D}
leads to a homogeneous system for two (complex) equations for the $A_i$, 
\be \label{M2}
\mathbf{M}(\nu_o, \omega, \Omega, c, \rho, m, R)   \left(
\begin{array}{c}
A_1\\ A_2
\end{array}
\right) =0.
\ee
The complex matrix $\mathbf{M}$ also depends on $\kappa_j$ through the dispersion relation. 
System \rr{M2} has a solution only when its determinant vanishes, explicitly when 
\be \label{detM}
\textrm{det} \mathbf{M} \equiv v_r(\kappa_1R) v_\phi(\kappa_2 R) - v_r(\kappa_2R) v_\phi (\kappa_1R) =0.
\ee
We thus parameterize $\kappa_1(\omega, \nu_o)$ and $\kappa_2(\omega, \nu_o)$, fix values for $\Omega, c, \rho, m$
and $R$, substitute into the real and imaginary parts of the secular equation \rr{detM} and 
solve for $\omega$ and $\nu_o$.

\begin{figure}
   \begin{subfigure}[t]{0.48\textwidth}
        \includegraphics[width=\linewidth]{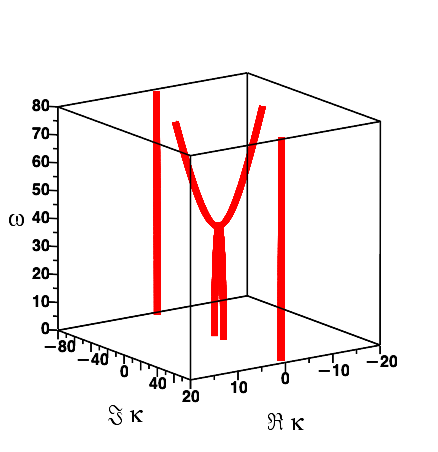} 
        \caption{$(c, \Omega, \nu_o, \rho) = (8, -20, 0.1, 1)$} \label{bif1}
    \end{subfigure}
    \begin{subfigure}[t]{0.52\textwidth}
        \includegraphics[width=\linewidth]{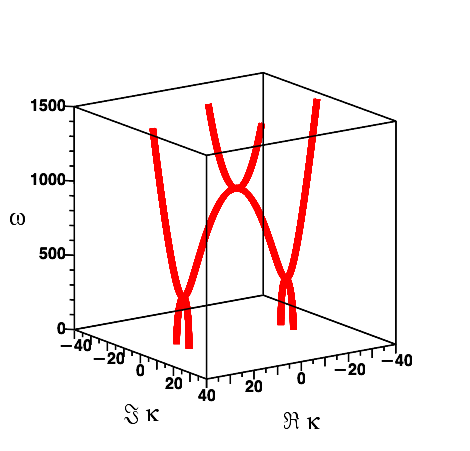} 
       \caption{$(c, \Omega, \nu_o, \rho) = (15, -500, 2, 1)$} \label{bif2}
    \end{subfigure}
    \caption{\label{bif}Real frequency $\omega$ versus the real and imaginary parts of 
    the eigenvalue $\kappa $ derived as solution of Eq. \rr{kappa2c}. 
    Both panels emphasize the presence of imaginary or complex values of $\kappa$ that cannot be captured
    by a plane-wave analysis of the momentum equations. In particular, the domain $\omega<2\Omega$
    is populated by imaginary or complex $\kappa$'s that may give rise to wall (evanescent wave) modes
    (the character of $\kappa$ is displayed in figure \ref{roots2D}). 
    Note that the indicated curves are symmetric with respect
    to the $\omega=0$ plane and continuously extend toward negative $\omega$ values. The parameters
    are given in arbitrary units. 
    } 
\end{figure}

\subsection{\label{sec: kappa2}Character of the eigenvalue $\kappa$}
Recent literature on two-dimensional compressible odd viscous liquids \citep{Tauber2019,Souslov2019}
has brought forward examples supporting the bulk-interface correspondence by establishing the presence of waves
propagating parallel to a boundary and increasing/decreasing exponentially with distance from it. 
The starting point for these studies is the dispersion relation $\omega = \omega (\mathbf{k})$ obtained from 
\rr{disp2} where $\mathbf{k}$ is a real wavevector. In this case dispersion curves exists only for $\omega>2\Omega$.
To obtain the exponential behaviour the wavenumber then
is set to be complex and the frequencies studied are those in the ``gap'', that is, with $\omega<2\Omega$.

We here adopt an opposite outlook, similar to the one followed in the literature
of rigidly-rotating liquids. This method emphasizes the derivation of a wavenumber $\kappa$, as in Eq. \rr{kappa2c}, that can 
be complex and is a function of an always real frequency $\omega$.

\begin{figure}
\vspace{5pt}
\begin{center}
\includegraphics[height=3.4in,width=5.6in]{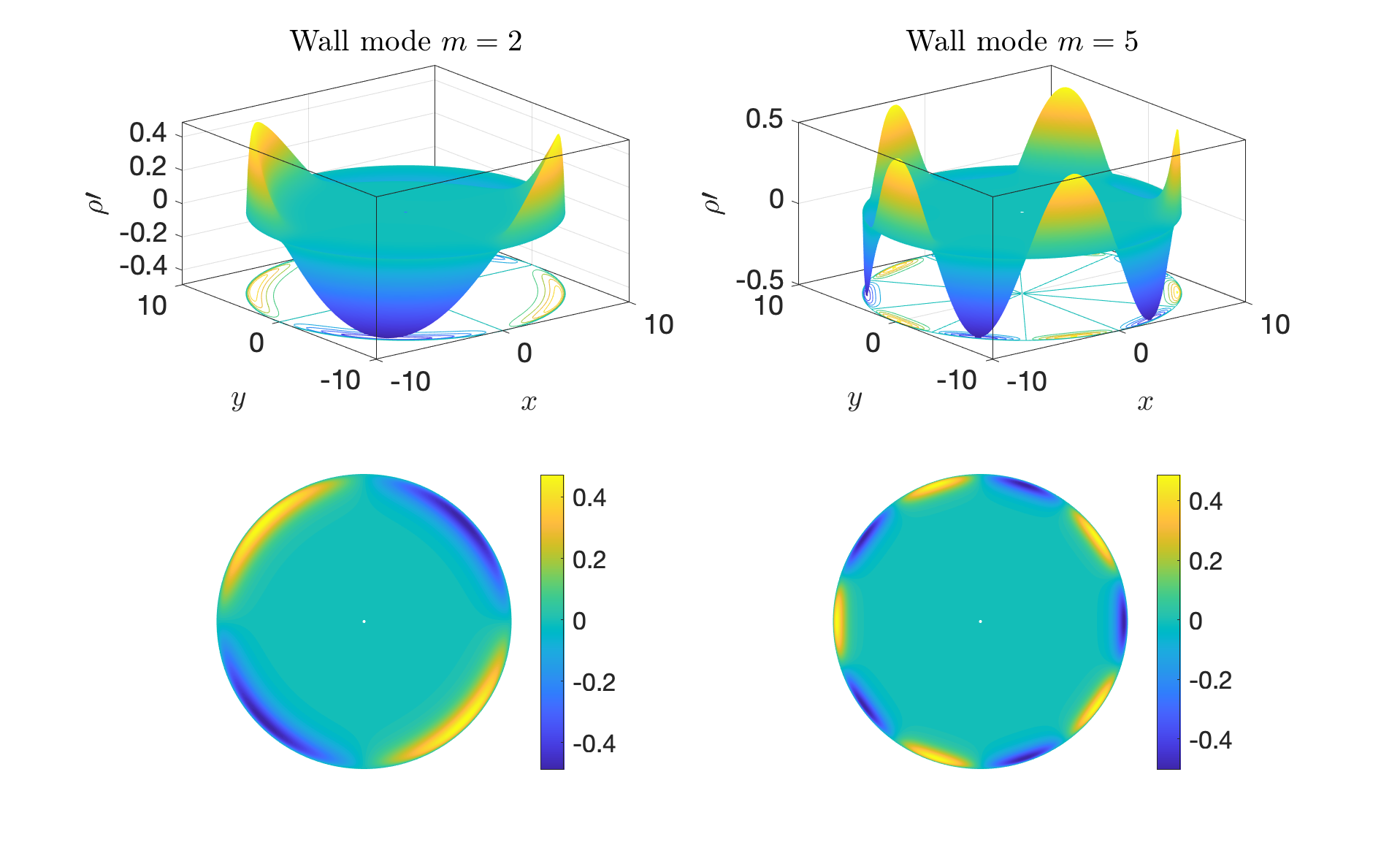}
\vspace{-10pt}
\end{center}
\caption{
\label{wall_modes2D} Density profiles and contours for the wall modes arising when the parameter $\kappa$ lies in the lower left of the diagram
 \ref{roots2D} (two imaginary pair $\kappa$'s) and thus the frequencies lie in the ``gap'' $(-2\Omega, 2\Omega)$. 
 Left column: $m=2$ mode, with $(\omega, \nu_o) = (0.4, 6.4)$ as solution of system \rr{detM} leading to $(\kappa_1, \kappa_2) = (-1.95i, -3.2i)$. Right column:
    $m=5$ mode, with $(\omega, \nu_o) = (1.5, 2.5)$ as solution of system \rr{detM} leading to $(\kappa_1, \kappa_2) = (-2.7i, -5.9i)$. In both cases $(c, \Omega, \rho_0, R) = (8, 20, 1,10)$ and thus both profiles precess in a prograde manner in the frame rotating with the liquid. Note the resemblance of the density profiles
 with the temperature distribution of rapidly-rotating (non-odd) Rayleigh-B\'enard convection in \citep[Fig. 6]{Goldstein1993}
and of the contour plots with those of \citet[Fig. 3 and S3]{Souslov2019}. Observing experimentally the precession
rate of patterns could, in principle, lead to the determination of the odd viscosity coefficient. Parameter and observable
units are arbitrary. 
 }
\vspace{-0pt}
\end{figure}

We display in figure  \ref{bif} the frequency $\omega$ vs. the real and imaginary parts of $\kappa$, drawn from \rr{kappa2c} which is to be compared with figure \ref{roots2D} and table \ref{tab: roots}. In panel \ref{bif1}, when  $\omega<2|\Omega| = 40$ ($\beta <0$), 
the four imaginary roots are clearly visible. When $\omega > 40 = 2|\Omega| $ ($\beta>0$), Eq. \rr{kappa2c} acquires two real and two imaginary $\kappa$ roots. 
In panel  \ref{bif2}, when $\omega<2|\Omega| = 1000$ ($\beta <0$), 
two distinct behaviours are visible. Those that make $\alpha^2+\beta$ in \rr{kappa2c} negative, giving 
rise to the aforementioned complex roots (located below the parabola of Fig. \ref{roots2D}), while
those that make $\alpha^2+\beta$ positive give rise to four real roots (located outside the parabola of Fig. \ref{roots2D}). When $\beta$ becomes positive ($\omega>2|\Omega| = 1000$), we obtain two
imaginary and two real roots for $\kappa$. 

The conclusion of this discussion is that values of the frequency $\omega<2\Omega $ exist when the wavenumber $\kappa$ is imaginary or complex, and this is a natural outcome of the present formulation. Note that both panels
in figure \ref{bif} symmetrically extend for negative values of the frequency.

\subsection{Basic observations}
From \rr{nsrc}-\rr{nszc} we can calculate the vorticity $\textrm{curl}\mathbf{v} = \frac{1}{r}\left[\frac{\partial (r v_\phi)}{\partial r} -\frac{\partial  v_r}{\partial \phi} \right] \hat{\mathbf{z}}$ of the two-dimensional odd viscous liquid, which, is found to be  proportional to the density $\rho'$:
\be \label{vortrho}
\textrm{curl}\mathbf{v} 
= (\kappa^2 \nu_o + 2\Omega) \frac{\rho'}{\rho}\hat{\mathbf{z}}.
\ee
This proportionality was also pointed-out in the numerical simulations of \citet[Fig. S2]{Souslov2019}. 
On account of the equivalence of the two and three-dimensional problem (to be discussed in section \ref{sec: equivalence}), this proportionality 
is justified on the basis of conservation of helicity. 
In addition, it 
is expected to persist, even when (the neglected here)
nonlinear terms are incorporated into the Navier-Stokes equations \citep{Pelz1985}. 
The conservation of helicity follows familiar lines and is thus relegated to the Supplementary Materials addendum.

%
%

\subsection{\label{sec: wallbody2}Wall and body modes in two-dimensional compressible odd viscous liquids}
\citet{Goldstein1993,Knobloch1994}, following the experiments of \citet{Ecke1992} in rapidly-rotating Rayleigh-B\'enard convection (of a non-odd liquid), showed theoretically the existence of two types of non-axisymmetric modes precessing in the rotating system: 
\emph{Wall modes} which peak near the sidewall and decay in the interior of the cylinder and \emph{body modes}
filling the whole cylinder and having their largest amplitudes close to the center rather than the sidewall. 
Both types are classified in table \ref{tab: roots}. 
We proceed by showing that the two-dimensional compressible odd viscous liquid under consideration
gives rise to similar wall and body modes which can 
be understood in the context of our formulation as evanescent and oscillatory inertial-like waves, respectively.

\begin{figure}
\vspace{5pt}
\begin{center}
\includegraphics[height=2.6in,width=2.8in]{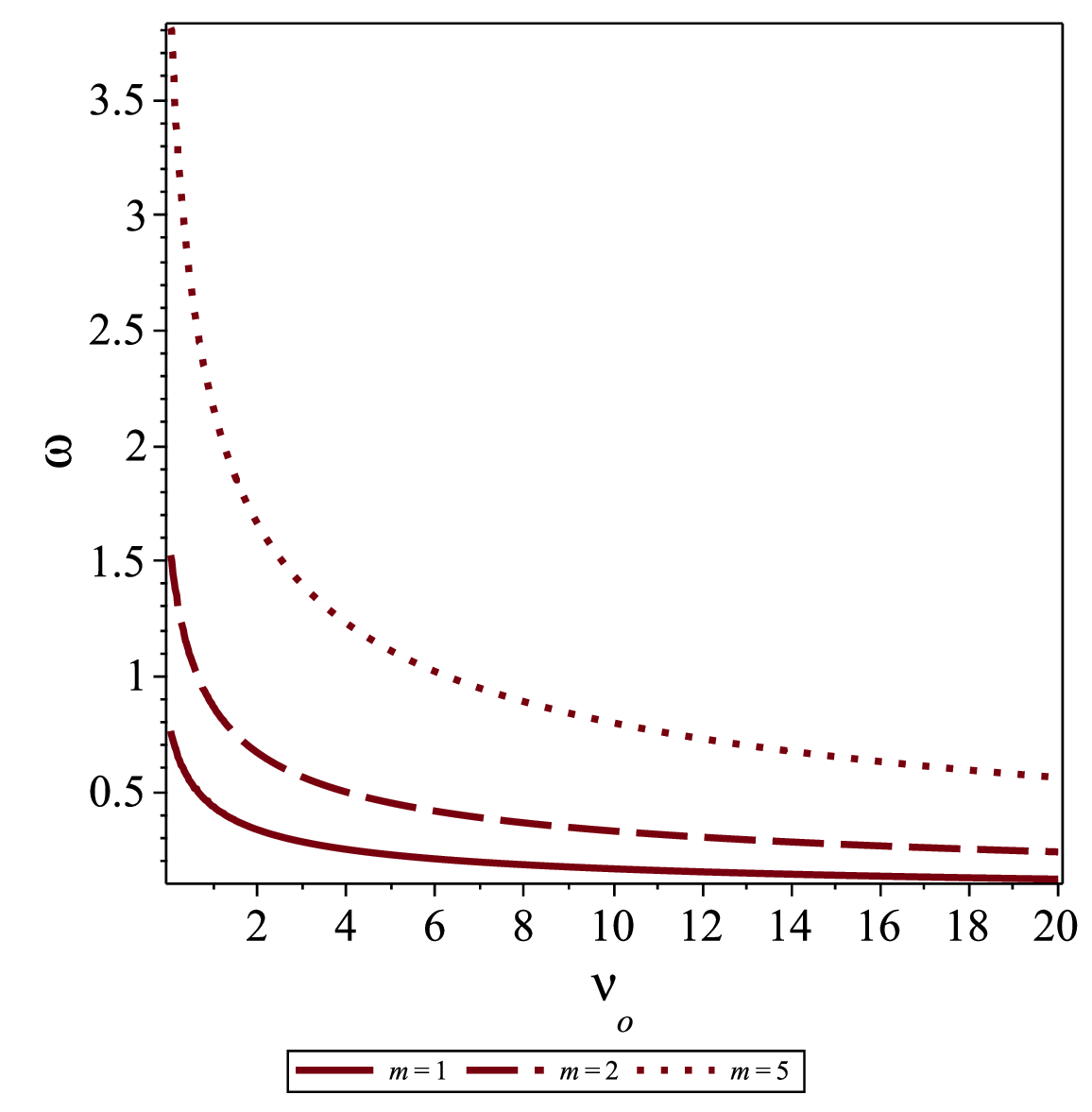}
\vspace{-0pt}
\end{center}
\caption{
\label{omega_nu0_bif}Admissible $(\nu_o, \omega)$ pairs, as solution of system \rr{detM}, giving rise to the wall modes displayed in Fig. \ref{wall_modes2D},
employing the latter figure's parameter values. Thus, observing experimentally the precession
rate of patterns $\omega$, it would be possible, in principle, to determine the largely unknown value of the odd viscosity coefficient $\nu_o$. Arbitrary units of the parameters were employed. 
}
\vspace{-0pt}
\end{figure}

In figure \ref{wall_modes2D} we display the density profiles (first row) for the \emph{wall} modes $m=2$ and $m=5$ arising
by solving the system \rr{detM} in a disk of radius $R=10$ rotating with angular velocity $\Omega = 20$ and 
$\omega>0$ in both cases (we employ arbitrary units). 
Thus, the profiles precess with frequency $\omega$ 
in a prograde manner in the frame rotating with the disk. Both profiles resemble those of the temperature
distribution in (non-odd) rapidly-rotating Rayleigh-B\'enard convection as they are displayed in Fig. 6 of \citet{Goldstein1993}. Note also the resemblance of the contour plots (second row) to the ones of \citet[Fig. 3 and S3]{Souslov2019}.

In figure \ref{omega_nu0_bif} we display admissible $(\nu_o, \omega)$ pairs, as solution of system \rr{detM}, giving rise to the \emph{wall} modes displayed in Fig. \ref{wall_modes2D}. Thus, all the corresponding $\kappa$'s arising
from the displayed parameter pairs are imaginary and will give rise to exponentially decaying velocity fields in the radial direction. 
Since the modes precess with frequency $\omega$ in the rotating frame, these modes will also propagate
parallel to the circular boundary of the disk. It is thus clear that observing experimentally the precession
rate of patterns could, in principle, lead to the determination of the odd viscosity coefficient $\nu_o$. 

We note that \citep{Favier2020,Knobloch2022} commented about the resemblance of wall modes in (non-odd) rapidly-rotating Rayleigh-B\'enard convection to odd viscosity-induced topological waves that appear near the boundary of a 
rotating two-dimensional odd viscous liquid \citep{Souslov2019}.

\begin{figure}
\vspace{5pt}
\begin{center}
\includegraphics[height=1.8in,width=5.6in]{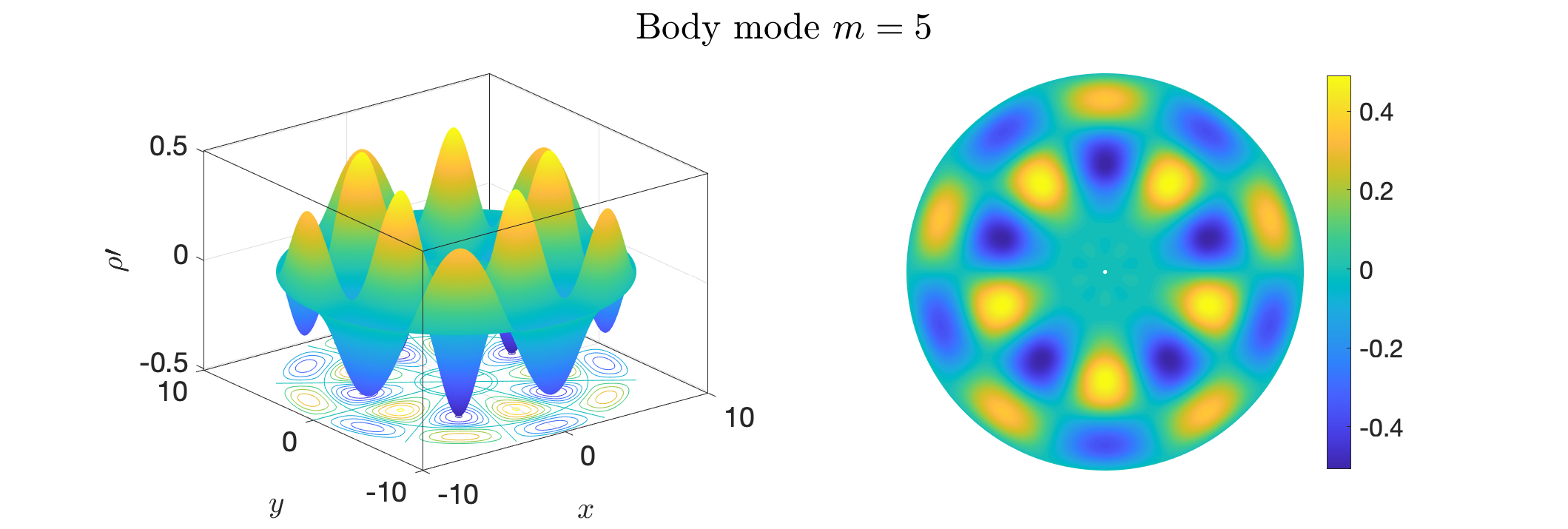}
\vspace{-0pt}
\end{center}
\caption{
\label{body_mode5_2Db}A body mode for $m=5$ with $(\omega, \nu_o) = (34.6, -2.8)$ as solution of system \rr{detM} leading to $(\kappa_1, \kappa_2) = (-4.2, 1.7)$, and same parameters as in Fig. \ref{wall_modes2D}. Thus the admissible $\kappa$'s are located in the lower right of Fig. \ref{roots2D}. The patterns precess in the rotating frame in a prograde manner. Note the resemblance of the density profiles
 with the temperature distribution of non-odd rapidly-rotating Rayleigh-B\'enard convection in \citep[Fig. 7]{Goldstein1993}. Observing experimentally the precession
rate of patterns could, in principle, lead to the determination of the odd viscosity coefficient. Units employed above
are arbitrary. }
\vspace{-0pt}
\end{figure}

In figure \ref{body_mode5_2Db} we display the density and contour profile of the \emph{body} mode $m=5$ arising
by solving the system \rr{detM} in a disk of radius $R=10$ rotating with angular velocity $\Omega = 20$. 
The profile precesses with frequency $\omega$ 
in a prograde manner in the frame rotating with the disk. The profile resembles somewhat the temperature
distribution in the (non-odd) rapidly-rotating Rayleigh-B\'enard convection as they are displayed in Fig. 7 of \citet{Goldstein1993}.

\section{\label{sec: inertial}Inertial-like waves in a three-dimensional rigidly-rotating incompressible odd viscous liquid}
The constitutive law of an odd viscous liquid in three dimensions is of the form 
\be\label{sigma}
\bm{\sigma}' = \bm{\sigma}_o' + \bm{\sigma}_4', 
\ee
where, in cylindrical polar coordinates $r, \phi, z$,
\be \label{sigma0}
\bm{\sigma}_o' = \eta_o 
\left(\begin{array}{ccc}
-\left(\partial_r v_\phi - \frac{1}{r}v_\phi + \frac{1}{r}\partial_\phi v_r \right) 
&  \partial_r v_r - \frac{1}{r}v_r - \frac{1}{r}\partial_\phi v_\phi & 0\\
\partial_r v_r - \frac{1}{r}v_r - \frac{1}{r}\partial_\phi v_\phi  &  \partial_r v_\phi - \frac{1}{r}v_\phi + \frac{1}{r}\partial_\phi v_r  & 0\\
0&0&0
\end{array}
\right),
\ee
and 
\be \label{sigma4rphi}
\bm{\sigma}_4' = \eta_4 
\left(\begin{array}{ccc}
0 & 0  & -(\frac{1}{r}\partial_\phi v_z + \partial_z v_\phi)\\
0 & 0  & \partial_r v_z + \partial_z v_r\\
 -(\frac{1}{r}\partial_\phi v_z + \partial_z v_\phi)&\partial_r v_z + \partial_z v_r &0
\end{array}
\right)
\ee
where $\eta_o $ and $\eta_4$ are the odd viscosity coefficients. Notations employed in the literature to denote
the odd viscosity coefficients appear in table \ref{tab: coefficient}.

\begin{figure}
\vspace{5pt}
\begin{center}
\includegraphics[height=2.5in,width=1.4in]{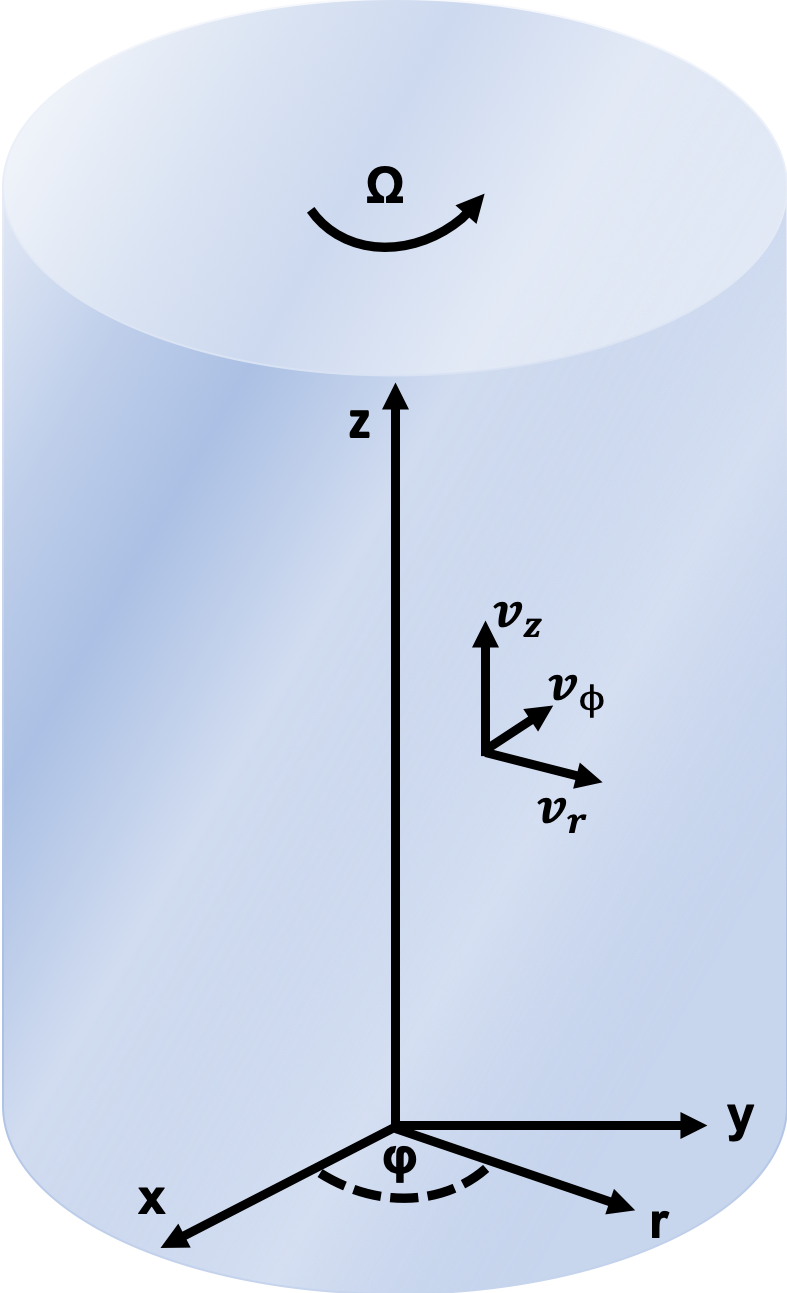}
\vspace{-0pt}
\end{center}
\caption{Three-dimensional odd viscous liquid rotating with angular velocity $\Omega$ about the 
$\hat{\mathbf{z}}$ axis. In cylindrical coordinates the velocity field is 
$\mathbf{v} = v_r \hat{\mathbf{r}} + v_\phi \hat{\bm{\phi}} + v_z \hat{\mathbf{z}}$ in the 
frame of reference rotating with the liquid. 
\label{cylinder}  }
\vspace{-0pt}
\end{figure}

Consider a three-dimensional odd viscous liquid rotating rigidly about the $\hat{\mathbf{z}}$ axis with angular velocity $\Omega$
(cf. Fig. \ref{cylinder}) and azimuthally-dependent fields
\be \label{vrhophit}
\left[v_r(r), v_\phi(r),v_z(r) , p'(r)\right] e^{i(kz+m\phi  -\omega t)}
\ee
in the frame of reference rotating with the liquid,
where the frequency $\omega$ and wave number $k$ along the axis are both real and $m$ is an integer. 
We assume that $k$ has already been fixed by suitable boundary conditions on the lids of the cylinder. 
We neglect the nonlinear terms by assuming small-amplitude motions. 
The 
Navier-Stokes equations take the form
\begin{align} \label{nsr2}
-i\omega v_r -2\Omega v_\phi& = -\frac{1}{\rho}\frac{\partial p'}{\partial r} 
- \nu_o\left[ \mathcal{L} v_\phi + \frac{2im}{r^2} v_r\right] + \nu_4 \left[k^2 v_\phi + \frac{mk}{r} v_z\right] , \\
-i\omega v_\phi + 2\Omega v_r & = -\frac{im}{r} \frac{p'}{\rho}
+
\nu_o\left[ \mathcal{L} v_r - \frac{2im}{r^2} v_\phi \right] + \nu_4 \left[-k^2 v_r + ik \frac{\partial v_z}{\partial r} \right] ,
\label{nsphi2}
\\
-i\omega v_z& = -\frac{i k}{\rho} \left[p' + \eta_4 \zeta \right], \label{nsz}
\end{align}
where $\mathcal{L}$ is the linear differential operator \rr{L}, $p'$ is the variable part of the pressure in the wave, $(\nu_o, \nu_4) \equiv( \eta_o, \eta_4)/\rho$ are the 
coefficients of kinematic odd viscosity, $\zeta$ is the $z$ component of the vorticity, $\zeta = \frac{1}{r} \left[ \frac{\partial}{\partial r} \left( r v_r\right)  - i m v_r \right]$, and the centrifugal acceleration has been combined into the
effective pressure $p'$, \citep{Greenspan1968}. 

The incompressibility condition becomes
\be \label{incomp2}
\frac{1}{r} \frac{\partial}{\partial r} \left( r v_r\right) +\frac{ im}{r} v_\phi+i kv_z= 0. 
\ee

\begin{table}
  \begin{center}
\def~{\hphantom{0}}
  \begin{tabular}{|c|c|c|c|}
\citep{Landau1981}& \citep{Khain2022}& \citep{Hulsman1970}& This paper\\
$\eta_3$ &$\eta_1^o$&$\eta_4$ & $\eta_o$\\
$\eta_4$&$\eta_2^o$&$\eta_5$ & $\eta_4$\\
  \end{tabular}
\caption{Conventions of odd viscosity coefficients that have appeared in the literature}
  \label{tab: coefficient}
  \end{center}
\end{table}

In Appendix \ref{sec: Poincare} we derive a single equation for the pressure $ \tilde{p} = p' + \eta_4 \zeta$
\be \label{PC3Da}
\left[\nabla_2^2 + \left(1 - \frac{(2\Omega - \mathcal{S})^2}{\omega^2}\right) \partial_z^2\right] \tilde{p} =0, 
\ee
where $\mathcal{S}$ is the linear operator 
 \be \label{S}
\mathcal{S} = (\nu_o-\nu_4)\nabla^2_2 + \nu_4\partial_z^2 ,
\ee
$\nabla^2_2$ is the two-dimensional (horizontal) Laplacian and we considered perturbations of the pressure $\sim \exp(-i\omega t)$ with real frequency $\omega$.
Clearly, when $\nu_o = \nu_4 =0$ Eq. \rr{PC3Da} reduces to the standard Poincar\'e-Cartan equation \rr{PCrot}
of non-odd rigidly-rotating liquids.  

Substituting 
\be
\tilde{p}(r,\phi) = J_m(\kappa r) e^{i(kz+m\phi -\omega t)}
\ee
into \rr{PC3Da}, where $J_m$ is the Bessel function of first kind, 
 leads to a quartic equation for the determination of wavenumber $\kappa$
\be\label{detrot}
-\kappa^{2}-k^{2} \left[1-\frac{\left(2 \Omega +\nu_4 \,k^{2}+\left(\nu_o -\nu_4 \right) \kappa^{2}\right)^{2}}{\omega^{2}}\right]
 =0.
\ee
The $\kappa$ solutions of \rr{detrot} are given by the simple expression
\be \label{kappa2}
 \kappa^2 = 
\frac{\alpha \pm \sqrt{\alpha^2 + \beta}}{2(\nu_o - \nu_4)^2 k^2},
\ee
with
\be \label{alphabeta}
\alpha = \omega^{2}+2 k^{2} \left(\nu_4 -\nu_o \right) (2\Omega + \nu_4 k^2), \quad 
\beta = 4k^4(\nu_o-\nu_4)^2\left[\omega^2 - (2\Omega+\nu_4k^2)^2\right]. 
\ee
Thus, the character of $\kappa$'s (real, imaginary or complex) is again described by figure \ref{roots2D}
and table \ref{tab: roots}. 

From the four values of $\kappa$ obtained in \rr{kappa2}, only two give rise to linearly independent solutions. Let 
$\kappa_1$ and $\kappa_2$ denote these $\kappa$ values. The fields can then be cast as
\be \label{3Dsol}
\left(
\begin{array}{c}
v_r \\ v_\phi
\end{array}
\right) = \sum_{j=1}^2
A_j\gamma_j \left(
\begin{array}{cc}
\delta_j  & 2\Omega \\
-2\Omega&  \delta_j
\end{array}
\right) \left(
\begin{array}{c}
\partial_r \\ \frac{im}{r}
\end{array}
\right)J_m(\kappa_j r), \quad v_z= \sum_{j=1}^2A_jJ_m(\kappa_j r),
\ee
the coefficients $\gamma_j$ and $\delta_j$ are 
\be \label{betagamma3D}
\gamma_j = \frac{\kappa_j^{2} (\nu_o-\nu_4) +\nu_4k^2 + 2 \Omega }{2 \Omega  \,\kappa_j^{2} \omega}, \quad 
\delta_j = \frac{-2 i \omega \Omega}{\kappa_j^{2} (\nu_o-\nu_4) +\nu_4k^2 + 2 \Omega }, \quad j=1,2,
\ee
determined by substituting the solution \rr{3Dsol} into the isochoric constraint \rr{incomp2} and into the $z$ component 
of the vorticity equation, and $A_j$ are complex constants to be determined by the boundary conditions. 

We impose no-slip boundary conditions at the sidewall of the cylinder $r=R$
\be 
v_r(r=R) = v_\phi(r=R) =0,
\ee
leading to a homogeneous system for two (complex) equations for the $A_i$, 
\be \label{M3}
\mathbf{M}(\nu_o, \nu_4,\omega, \Omega, k, m, R)   \left(
\begin{array}{c}
A_1\\ A_2
\end{array}
\right) =0.
\ee
System \rr{M3} has a solution only when its determinant 
\be \label{detM3}
\textrm{det} \mathbf{M} \equiv v_r(\kappa_1R) v_\phi(\kappa_2 R) - v_r(\kappa_2R) v_\phi (\kappa_1R) =0,
\ee
vanishes.
We thus parameterize $\kappa_1(\omega, \nu_o, \nu_4)$ and $\kappa_2(\omega, \nu_o,\nu_4)$, fix values for $\Omega, k, R,m$, substitute into the real and imaginary parts of \rr{detM3} and 
solve for $\omega$ and $\nu_o$ (see the discussion in section \ref{sec: mixed3} on how $\nu_4$ is chosen).

\begin{figure}
   \begin{subfigure}[t]{0.5\textwidth}
        \includegraphics[width=\linewidth]{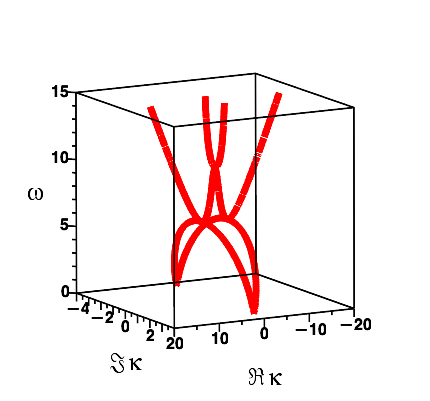} 
        \caption{$(\Omega, \nu_o, k) = (5, 1, 1)$} \label{bif31}
    \end{subfigure}
    \begin{subfigure}[t]{0.5\textwidth}
        \includegraphics[width=\linewidth]{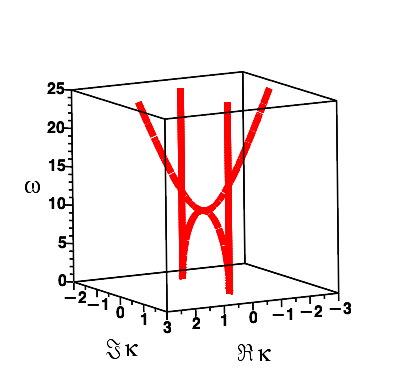} 
       \caption{$(\Omega, \nu_o, k) = (5, 10, 1)$} \label{bif32}
    \end{subfigure}
    \caption{\label{bif3}Real frequency $\omega$ versus the real and imaginary parts of 
    the eigenvalue $\kappa $ derived as solution of Eq. \rr{kappa2}. 
    Both panels emphasize the presence of imaginary or complex values of $\kappa$ that cannot be captured
    by a plane-wave analysis of the momentum equations. In particular the domain $\omega<2\Omega$
    is populated by imaginary or complex $\kappa$'s that exclusively give rise to wall (evanescent wave) modes. 
    Note that the indicated curves are symmetric with respect
    to the $\omega=0$ plane. Units employed above
are arbitrary.
    } 
\end{figure}

\subsection{Character of the $\kappa$ eigenvalues}
A non-odd rotating liquid has $\kappa$ solutions satisfying
$$
\kappa = k\sqrt{\frac{4\Omega^2}{\omega^2} -1} \quad \textrm{(non-odd rotating liquid)}
$$
(obtained from the Poincar\'e-Cartan Eq. \rr{PCrot}). 
Thus, when $\omega <2\Omega$, $\kappa$ is real and 
the solutions are oscillatory Bessel functions. When however $\omega >2\Omega$, there are two
imaginary $\kappa$'s and the fields are modified (exponentially increasing) Bessel functions. 

We display in figure  \ref{bif3} the frequency $\omega$ vs. the real and imaginary parts of $\kappa$, drawn from \rr{kappa2} which is to be compared with figure \ref{roots2D} and table \ref{tab: roots}. 

In panel \ref{bif31} we display the bifurcation diagram
    for the set of values $(\Omega, \nu_o, k) = (5, 1, 1)$ (arbitrary units). 
    There are two complex conjugate roots up to $\omega = 6$, where the radical in \rr{kappa2} changes sign. For $6<\omega<10 = 2\Omega$
    there are four real roots and beyond this, two imaginary and two real roots. 
In panel \ref{bif32} we employ the alternative set of values  
    $(\Omega, \nu_o, k) = (5, 10, 1)$ (arbitrary units). 
    There are 
    four imaginary $\kappa$ roots up to $\omega = 10 = 2\Omega$. Beyond this 
    there are two imaginary and two real roots.   
%
%

\subsection{\label{sec: equivalence}Formal equivalence of the two- and three-dimensional problems}
Setting $\nu_4=0$ in the equations of motion shows that they are equivalent to their two-dimensional 
counterparts by effecting the correspondence
\be
p' = c^2 \rho' \textrm{  (2D)}, \quad p' = \frac{\rho \omega}{k} v_z   \textrm{  (3D)}.
\ee
The pressure in the two and three-dimensional cases satisfies $p' = \frac{\rho c^2}{i\omega} \textrm{div}_2 \mathbf{v}$, 
and $p' =- \frac{\rho \omega}{ik^2} \textrm{div}_2 \mathbf{v}$, respectively, where $\textrm{div}_2 \mathbf{v} =\frac{1}{r} \left[\partial_r (rv_r) + im v_\phi \right]$. Thus, the two problems are identical inasmuch as the respective dispersion relations
are taken into account.

Likewise, the three-dimensional problem with $\nu_4 \neq 0$ can be recovered from the three-dimensional problem with
$\nu_4 =0$ by performing the substitution 
\be \label{nu0tonu4}
\nu_o \rightarrow \nu_o -\nu_4, \quad \textrm{and} \quad 2\Omega \rightarrow 2\Omega + \nu_4 k^2.
\ee
Thus, the role of $\nu_4$ is to renormalize both the angular velocity $\Omega$ and the odd viscosity coefficient
$\nu_o$. 

The question arises, if the two- and three-dimensional problems are mathematically equivalent, where do they 
differ? In the Appendix we show that the role of $\nu_4$ is to alter the direction of propagation of data along
characteristics. For instance, when $\nu_4$ is zero, characteristics are parallel to the $z$ axis, giving rise to 
a Taylor column, while when $\nu_4$ is non-zero they become oblique to the center axis.

\begin{figure}
\vspace{5pt}
\begin{center}
\includegraphics[height=3.4in,width=5.6in]{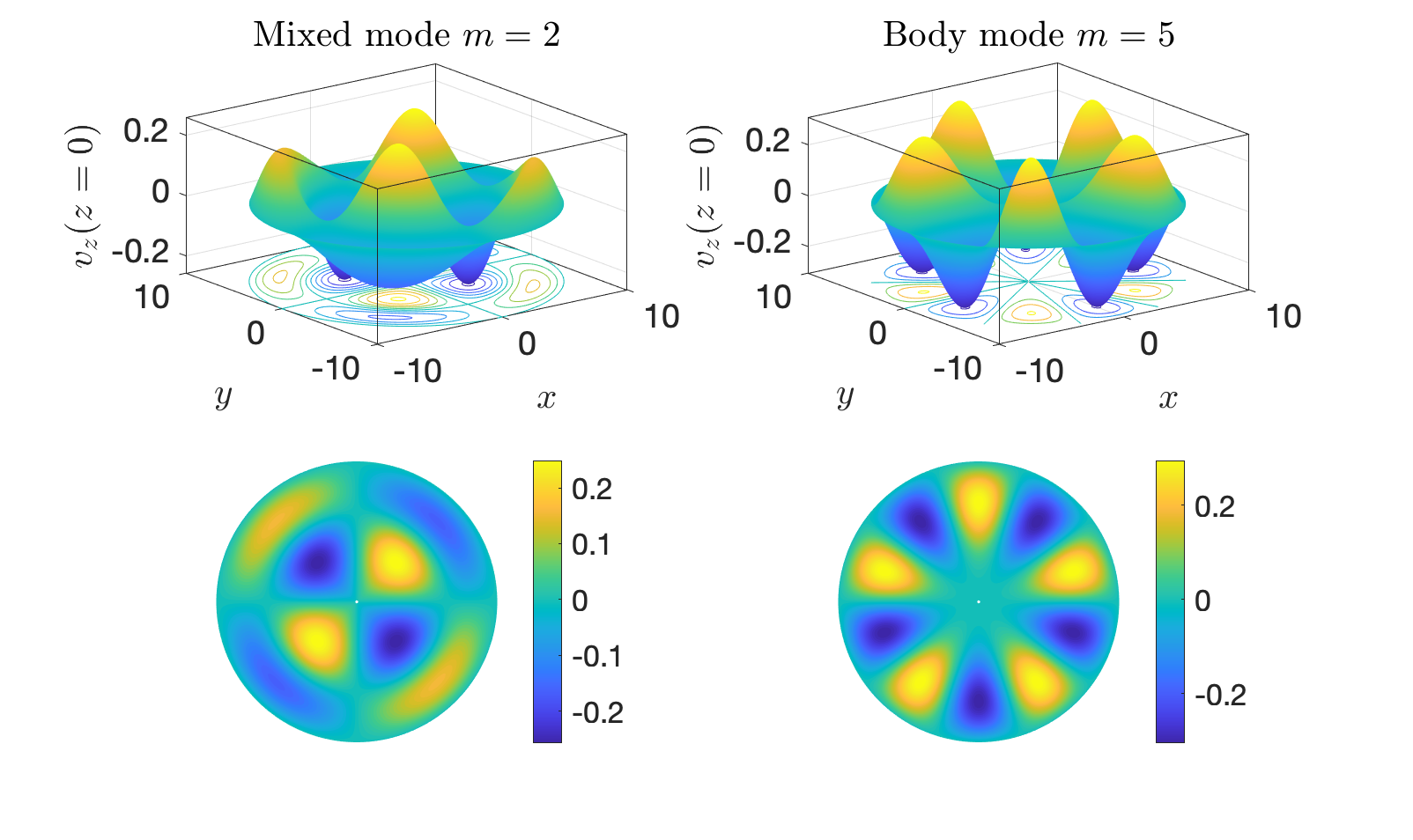}
\vspace{-10pt}
\end{center}
\caption{
\label{mixed_body_mode5_3D}Precessing axial velocity (or pressure) profiles and contours for the modes arising when the parameter $\kappa$ lies in the upper part (two real two imaginary) and lower right section of the diagram
 \ref{roots2D} (four real $\kappa$'s), respectively. 
 Left column: $m=2$ mixed mode, with $(\omega, \nu_o) = (-0.9, -3.3)$ as solution of system \rr{detM} leading to $(\kappa_1, \kappa_2) = (-0.97, -0.5i)$. Right column:
    $m=5$ body mode, with $(\omega, \nu_o) = (0.3,-3 )$ as solution of system \rr{detM} leading to $(\kappa_1, \kappa_2) = (-0.75, 0.3)$. In both cases $(k, \Omega, R) = (1, 1, 10)$ in arbitrary units. The odd visccosity coefficients were chosen to satisfy
    $\nu_o=2\nu_4$, as explained in \rr{eta02eta4}. Observing experimentally the precession
rate of patterns could, in principle, lead to the determination of the odd viscosity coefficients $\nu_o$ and $\nu_4$. Units employed above
are arbitrary.
 }
\vspace{-0pt}
\end{figure}

\subsection{\label{sec: mixed3}Mixed and body modes in a three-dimensional odd viscous liquid}
In section \ref{sec: wallbody2} we illustrated the two-dimensional theory by deriving the density profiles when
the wavenumbers $\kappa$ were all imaginary or all real, thus giving rise to wall and body modes defined in table \ref{tab: roots}, respectively as these
are displayed in figures 
\ref{wall_modes2D} and \ref{body_mode5_2Db}, respectively. There is a third category however, as this is displayed
by the character of wavenumbers $\kappa$ in the upper part of the diagram \ref{roots2D}, where there are concurrent real and imaginary admissible $\kappa's$ as solution of equation \rr{kappa2c}. Here we will provide one such example, 
which is displayed in the left collumn of Fig. \ref{mixed_body_mode5_3D}.

In three dimensions there are two odd viscosity coefficients and a choice has to be made regarding the solution
of secular equation \rr{detM3} (two equations for the determination of $\omega, \nu_o$ and $\nu_4$. We thus
follow \citep{Markovich2021,Khain2022} who consider the combination 
\be \label{eta02eta4}
\eta_o = 2\eta_4
\ee
as representing an experimentally-verified case (for polyatomic gases, \citep{Hulsman1970}). See the discussion
in \citep[\S8.3]{kirkinis2023axial} for some consequences of making this choice. 

In figure \ref{mixed_body_mode5_3D} we display the axial velocity profiles (first row) for the \emph{mixed} mode $m=2$ and body mode $m=5$ arising
by solving the system \rr{detM3} in a cylinder of radius $R=10$ rotating with angular velocity $\Omega = 1$, $k=1$, employing arbitrary units and enforcing the combination \rr{eta02eta4}. Plotting the velocity component $v_z$ is 
equivalent to plotting the pressure $\tilde{p} = p' + \eta_4 \zeta$ on account of the connexion \rr{nsz}. 
Note that the non-vanishing $\nu_4$ has renormalized the cylinder angular velocity $\Omega$ according to 
\rr{nu0tonu4}, making $\beta = 8.3>0$ in \rr{alphabeta}. 

As in the two-dimensional case, the outcome of this formulation is to give some means of determining the odd viscosity coefficients by observing experimentally the precession
rate of the patterns. 

%

\subsection{\label{sec: evanescent}Evanescent waves in the experiments of \citet{Nosan2021}}
Consider an annular cylinder of inner and outer radii $R_1$ and $R_2$, respectively, where the 
inner boundary is being radially displaced harmonically with frequency $\omega$. The liquid is 
rotating as a whole
with angular velocity $\Omega$ about the central axis. This then
is a system with the geometry employed in the recent experiments of \citep{Nosan2021}
which showed that a (non-odd) rigidly-rotating three-dimensional incompressible liquid 
gives rise to evanescent waves at the cross-over frequency $\omega \rightarrow 2\Omega$. 
Setting $\omega = 2\Omega$ ($\beta=0$ in \rr{kappa2}), 
we obtain two imaginary roots $\kappa=\pm i\tilde{\kappa}$, for real $\tilde{\kappa}$ from \rr{kappa2} if $0<\Omega <\Omega_o (\equiv \nu_ok^2)$ and the solution reads
\be 
v_r = AI_1(\tilde{\kappa} r) + BK_1(\tilde{\kappa} r),
\ee
where by $\tilde{\kappa}$ in this section only we denote the imaginary part of the roots of \rr{kappa2}. Thus,
\be
\tilde{\kappa} = 2\sqrt{\frac{\Omega}{\nu_o} \left( 1 - \frac{\Omega}{\Omega_o}\right)}, \quad \Omega_o = \nu_ok^2>\Omega,
\ee
where, on account of the discussion in the previous paragraph we've set $\eta_4\equiv 0$. 
Let 
\be \label{eta}
\eta(z,t) = \hat{\eta} e^{ i (kz-\omega t)}
\ee
be the \emph{radial} displacement of the inner boundary at $r=R_1$ with complex $\hat{\eta}$ which will excite inertial-like waves
of oscillatory or evanescent character. Thus, the radial velocity satisfies 
\be \label{vrR1}
v_r(R_1, z, t) = \partial_t \eta = -i\omega \eta, \quad v_r(R_2, z,t) =0.
\ee

\begin{figure}
\vspace{-5pt}
\begin{center}
\includegraphics[height=2.5in,width=3.5in]{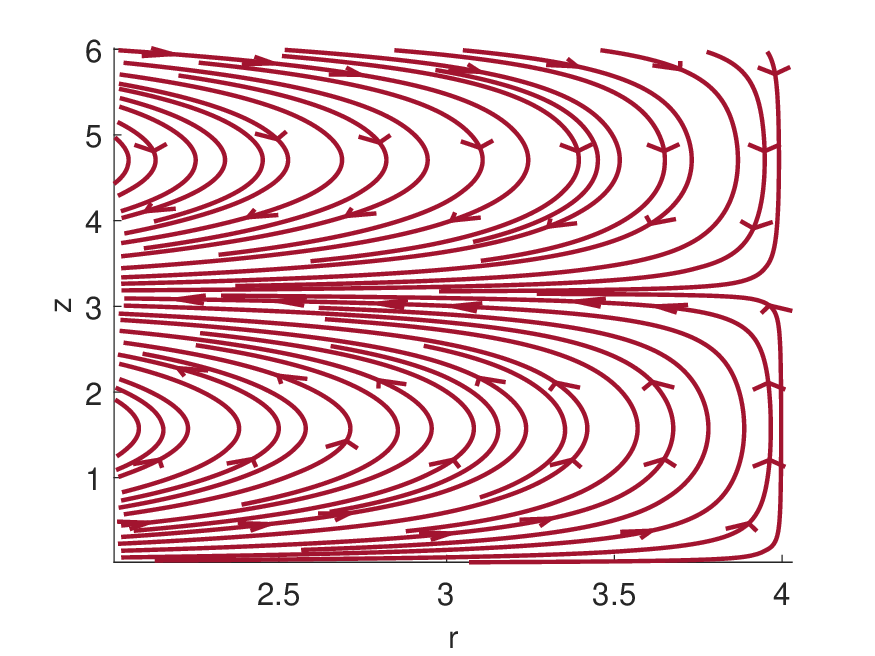}
\vspace{-0pt}
\end{center}
\caption{
Instantaneous streamlines in the $r-z$ plane of the velocity field \rr{vrev} - \rr{vzev}, representing a forced harmonic wave propagating in the $z-$ direction according to \rr{eta}. 
The liquid is confined between the forced inner cylinder at $r=R_1=2 + \Real \eta$ and the immobile external cylinder
at $r=R_2=4$ (in arbitrary units). 
\label{psi_zr2}  }
\vspace{-0pt}
\end{figure}
The boundary conditions \rr{vrR1} lead to the requirements 
$
v_r(R_1) = AI_1(\tilde{\kappa} R_1) + BK_1(\tilde{\kappa} R_1) = -i\omega \hat{\eta}$ and $
v_r(R_2) = AI_1(\tilde{\kappa} R_2) + BK_1(\tilde{\kappa} R_2) =0
$
from which we obtain
\be
A = -i\omega \hat{\eta} K_1(\tilde{\kappa} R_2)J^{-1},\; B = - A \frac{I_1(\tilde{\kappa} R_2) }{K_1(\tilde{\kappa} R_2) }
,\;
J = \left|
\begin{array}{cc}
I_1(\tilde{\kappa} R_1) & I_1(\tilde{\kappa} R_2)\\
K_1(\tilde{\kappa} R_1) & K_1(\tilde{\kappa} R_2)
\end{array} \right|. 
\ee
The solution simplifies somewhat if we cast $A$ in polar form $A = a e^{i\theta}$ for real $a, \theta$. 
Then, 
\be
a = \omega |\hat{\eta}|K_1(\tilde{\kappa} R_2)J^{-1},\quad \theta = -\arctan \frac{\textrm{Re}\hat{\eta}}{\textrm{Im}\hat{\eta}},
\ee
where $ |\hat{\eta}| = \sqrt{(\textrm{Re}\hat{\eta})^2 +(\textrm{Im}\hat{\eta})^2 }$ and 
$\textrm{Re}\hat{\eta}$ and $\textrm{Im}\hat{\eta}$ are the real and imaginary parts of the
complex number $\hat{\eta}$. 
We thus obtain, 
\begin{align} \label{vrev}
v_r &= \omega |\hat{\eta}|J^{-1}
\left[I_1(\tilde{\kappa} r) K_1(\tilde{\kappa} R_2)-I_1(\tilde{\kappa} R_2)  K_1(\tilde{\kappa} r)\right]\cos(kz - \omega t + \theta), \\
v_\phi &=(2\Omega - \nu_o\tilde{\kappa}^2) |\hat{\eta}|J^{-1}  \label{vphiev}
\left[I_1(\tilde{\kappa} r) K_1(\tilde{\kappa} R_2)-I_1(\tilde{\kappa} R_2)  K_1(\tilde{\kappa} r)\right]\sin(kz - \omega t + \theta), \\
v_z &=-\omega |\hat{\eta}|J^{-1}\frac{\tilde{\kappa}}{k}
\left[I_0(\tilde{\kappa} r) K_1(\tilde{\kappa} R_2)+I_1(\tilde{\kappa} R_2)  K_0(\tilde{\kappa} r)\right]\sin(kz - \omega t + \theta).
\label{vzev}
\end{align}
Note the positive sign of $K_0(\tilde{\kappa} r)$ in \rr{vzev} obtained because $K_n$ satisfies different derivative relations
to $I_n$. The solution \rr{vrev}-\rr{vzev} for evanescent waves in an odd viscous liquid is formally
analogous to the one obtained by \citet{Nosan2021}. We display the instantaneous resultant streamlines of system \rr{vrev}-\rr{vzev} 
in an $r-z$ slice of the cylinder in Fig. \ref{psi_zr2}. 

One could attempt to determine the value of odd viscosity with the experimental apparatus of \citet{Nosan2021} by measuring the planar velocity  $({v}_r, {v}_\phi)$ at fixed radial locations and different elevations, and averaging the result over a period of oscillation $2\pi/k$. 
These measurements could then be substituted into the expression 
(determined from equations \rr{vrev} and \rr{vphiev})
\be \label{ellipse}
\frac{{v}_r^2}{\left[ 2\Omega (J\tilde{\kappa})^{-1} |\hat{\eta}| \partial_rI(r) \right]^2 } + \frac{{v}_\phi^2}{\left[ (2\Omega - \nu_o\tilde{\kappa}^2) (J\tilde{\kappa})^{-1} |\hat{\eta}| \partial_r I(r)\right]^2 }= 1,
\ee
which only depends on the radial position of the measurement through the expression $\partial_r I(r)$, where
\be
I(r) = \left[I_0(\tilde{\kappa} r) K_1(\tilde{\kappa} R_2)+I_1(\tilde{\kappa} R_2)  K_0(\tilde{\kappa} r)\right]. 
\ee
Fixing the $r$ location, Eq. \rr{ellipse} then constitutes one algebraic equation to determine $\nu_o$.

The fields \rr{vrev} and \rr{vzev} can also be employed to determine the paths of fluid particles in the wave. 
Let $(r_0,z_0$) and $(r,z)$ be the equilibrium position and coordinates, respectively, of a moving fluid particle. 
Let $dr/dt = v_r$ and $dz/dt = v_z$ be the velocity of a fluid particle in the $r-z$ plane and consider small
oscillations away from the equilibrium position. Integrating with respect to time gives
the trajectories of the fluid particles which are the ellipses 
\be
\frac{(r-r_0)^2}{[k\partial_r I(r)]^2} + \frac{(z-z_0)^2}{I^2(r)} = \left[ \tilde{\kappa}  |\hat{\eta}|  J^{-1} \right]^2. 
\ee
Thus, observing the trajectories of suspended particles should, in principle, provide an alternative way of determining the value of the odd viscosity coefficient $\nu_o$.

Another related question regards the effect shear viscosity has on the motion of an odd viscous liquid. 
With respect to the present axisymmetric geometry, the kinematic shear viscosity $\nu$ enters into the Navier-Stokes equations just by 
adding the terms $\nu (\mathcal{L} -k^2) v_r$, $\nu (\mathcal{L} -k^2) v_\phi$ and $\nu (\mathcal{L} -k^2 +r^{-2}) v_z$,  
to the right hand side of \rr{nsr2}, \rr{nsphi2} and \rr{nsz}, respectively, (setting $m=\nu_4=0$) everything else remaining the same. 
Defining the dimensionless frequencies
\be
\xi = \frac{\nu k^2}{\omega}, \quad \textrm{and} \quad \xi_o = \frac{\nu_o k^2}{\omega},
\ee
and assuming velocity fields of modified Bessel function type $I_n(k\hat{\kappa r})$, leads the dimensionless wavenumber $\hat{\kappa}$ to satisfy
\be \label{kappazeta}
\xi^{2} \hat{\kappa}^{6}-\left(\xi_{o}^{2}+3 \xi^{2}-2 \,\mathrm{i} \xi\right) \hat{\kappa}^{4}+\left(\frac{4 \Omega  \xi_{o}}{\omega}-1+3 \xi^{2}-4 \,\mathrm{i} \xi\right) \hat{\kappa}^{2}-\xi^{2}+2 \,\mathrm{i} \xi +1 - \left(\frac{2\Omega}{\omega} \right)^2 =0.
\ee
Letting $\xi_o \equiv 0$ \rr{kappazeta} recovers the evanescent disturbance wavenumber equation (5.16) of \citet{Nosan2021}. On the other hand, setting $\xi=0$ we recover \rr{detrot}. Following the discussion of 
\citep{Nosan2021}, we can study the effects of shear viscosity perturbatively, more specifically, by considering $\xi$ as a perturbing parameter in Eq. \rr{kappazeta} and expanding $\hat{\kappa}$ in powers of $\xi$. The wavenumber $\hat{\kappa}$ correct to first order in $\xi$ is   
\be \label{kappahatzeta}
\hat{\kappa} \sim \frac{\sqrt{2 \xi_o-1}}{\xi_o} + \xi \frac{\mathrm{i} \left(\xi_o-1\right)^4}{\left(2 \xi_o-1\right)^{\frac{3}{2}} \xi_o^{3}}. 
\ee
Following the program established by \citet{Nosan2021}, the effect of shear viscosity $ \nu$ on the flow can now be determined by substituting \rr{kappahatzeta} into
\rr{vrev}-\rr{vzev} and expanding the modified Bessel functions $I_n(k\hat{\kappa}r)$ and $K_n(k\hat{\kappa}r)$ with respect to the perturbing parameter $\xi$. In the experiments of \citet{Nosan2021} (carried-out with a non-odd viscous liquid), this parameter was equal to 
$0.008$ which led to only modest viscous correction for $\omega \sim 2\Omega$ (with only one resonant exception). 
It is clear that these same conclusions are valid in the present case.

\section{Conclusion}
The main result of this paper is the derivation of precessing wall, body and mixed modes, as these are depicted in 
Fig. \ref{body_wall_cartoon}
and defined in 
table \ref{tab: roots}, in an odd viscous liquid as a consequence
of the complex or real character, respectively, of a planar wavenumber $\kappa$. The wall modes are 
evanescent waves and 
resemble the wall modes obtained in (non-odd) rotating Rayleigh-B\'enard convection \citep{Goldstein1993,Knobloch1994}, although the latter are a consequence of thermal
forcing in a liquid endowed with shear viscosity. They can also be understood as equatorial  
\citep{Tauber2019} and topological waves \citep{Souslov2019}
(that is, waves that propagate parallel to a boundary and decay away from it exponentially). 
That these modes
should be present in (the dispersive) odd viscous liquid system was also commented by \citet{Favier2020,Knobloch2022}. 

The analysis is essentially exact (subject to solving two transcendental equations), and gives rise to a parameter space which could be employed, in principle,
to determine the value of the odd viscosity coefficients. In a real system, shear viscosity will
be present and it will have to be taken into account. However, by itself, the presence of shear viscosity 
will always lead an initially forced system to decay in finite time.   
A meaningful system is one where some form of persistent forcing is always present. For instance in the experiments
of \citet{Soni2019}, which are the only ones where the odd viscosity coefficient has been determined in a liquid, the forcing was provided by a rotating magnetic field. Here, we only discuss some general behaviour of the dispersive system
without shear viscosity, with the exception of the \citep{Nosan2021} experiments in section \ref{sec: evanescent}. Effects of shear viscosity are expected to follow  familiar lines
of rotating liquids \citep{Chandrasekhar1961}. 


In the main body of this article we were concerned with the establishment of fluid flow behaviour that is 
non-axisymmetric, $m\neq 0$. 
In the Supplementary Materials addendum we provide a detailed discussion of the $m=0$ case (axisymmetric)
which includes elements of plane polarized waves and the conservation of helicity, following arguments analogous to those of  \citep{kirkinis2023axial}.

Rotating and stratified Boussinesq flow can be decomposed in parameter regimes depending
on the combination of strengths of these two effects \citep{Embid1998, Whitehead2014}. 
The typical characteristic of these systems is the existence of a two-dimensional slow manifold towards 
which
energy is being transferred due to fluctuations. Such a two-dimensional manifold is 
expected to exist in the case of a rigidly-rotating odd viscous liquid although the particle paths
of odd viscous liquids are circular while those of internal gravity waves are rectilinear \citep{Maas2001}. 
%
\\\\
\noindent
\textbf{Acknowledgments}\\ 
This research was supported by the US Department of Energy, Office of Science, Basic Energy Sciences under Award No. DE-FG02-08ER46539. The authors are grateful to the anonymous referees for their comments and suggestions that
significantly improved the manuscript. 
\\\\
\textbf{Declaration of Interests}\\ The authors report no conflict of interest.

\appendix
\section{Poincar\'e-Cartan equation for a (non-odd) rotating inviscid liquid}
For a (non-odd) liquid, rigidly-rotating about the $z$-axis with angular velocity $\Omega$, the Poincar\'e-Cartan equation \cite[\S2.6]{Greenspan1968}, is a second order partial differential equation satisfied by the pressure 
\be \label{PCrot}
\left[\nabla_2^2 + \left(1 - \frac{4\Omega^2}{\omega^2}\right) \partial_z^2\right] p' =0,
\ee
see also, \cite[\S 14]{Landau1987}. When $\omega<2\Omega$ this equation becomes
hyperbolic while, in the opposite case it becomes elliptic, see \cite[\S 12.6]{Whitham1974}. 
In the hyperbolic case the data propagate on characteristics lying on a cone making an angle $2\theta$
with the $z$-axis, where $\sin\theta= \omega/(2\Omega)$. Unsteady motions give rise to inertial waves, 
that is, plane-polarized waves with $
p' \sim e^{i(\mathbf{k}\cdot \mathbf{r} - \omega t)} 
$
($\mathbf{k} = (k_x,k_y,k_z)$ is the wavenumber whose magnitude is $k=|\mathbf{k}|$, and $\omega$ is the frequency of the inertial wave),
and to a dispersion relation 
$\omega = 2\Omega k_z/k$. 
On the other hand, steady motions $(\omega\equiv 0)$ lead to the requirement that 
$\partial^2_z p =0$. Characteristics are then straight lines parallel to the $z$-axis. Thus, data that emanate 
from a body moving slowly along the axis of rotation (the $z$-axis), propagate in straight lines ahead and behind the obstacle
forming Taylor columns \citep{Moore1968,Maxworthy1970,Bush1994} as long as the group velocity exceeds the speed of the slowly-moving body. 

\section{\label{sec: Poincare}Poincar\'e-Cartan equation of three-dimensional odd viscous liquids and its
classification}
In this section we will derive the Poincar\'e-Cartan equation for a rigidly-rotating odd viscous liquid, the analogue of \rr{PCrot}, which is then employed in equations \rr{detrot} and \rr{kappa2} to determine the admissible planar wavenumber values. An equivalence between the two and three-dimensional problems was established in 
section \ref{sec: equivalence}. Here, we proceed by showing that their difference lies on how data propagate along characteristic rays determined by the values taken over by the odd viscosity coefficients $\nu_o$ and $\nu_4$. 
This requires the consideration of the higher order terms of the governing PDEs only (the characteristic form), and in doing so we adopt the analysis of 
\citet[Ch. III, \S 2.3-2.6]{Courant1962}.

To this end, the Navier-Stokes equations take the form
(when the odd viscous liquid rotates with angular velocity $\Omega$), 
\be \label{NS2}
\frac{\partial u}{\partial t} = - \frac{1}{\rho}\frac{\partial \tilde{p}}{\partial x} + (2\Omega -\mathcal{S} )v, \quad \frac{\partial v}{\partial t} = -\frac{1}{\rho}\frac{\partial \tilde{p}}{\partial y} -(2\Omega- \mathcal{S} )u, \quad \frac{\partial w}{\partial t} = -\frac{\partial \tilde{p}}{\partial z}, 
\ee
where $\mathcal{S}$ is the second order differential operator defined in \rr{S} and $\tilde{p} = p'  + \eta_4 \zeta$, 
with $\zeta$ being the $z$-component of vorticity. 
In what follows we drop the superposed tilde on the pressure.

To derive the Poincar\'e-Cartan equation we follow the non-odd, rotating liquid procedure, cf. \cite[\S 14]{Landau1987}. Differentiate Eq. \rr{NS2} with respect to $x, y$ and $z$ and add to obtain
$-\nabla^2 p + \rho(2\Omega - \mathcal{S})(\partial_x v- \partial_y u) = 0$ where we employed the incompressibility
condition. Differentiating with respect to time and using \rr{NS2} again leads to 
$\partial_t \nabla^2p = - \rho (2\Omega - \mathcal{S})^2 \partial w/\partial z$ by employing the incompressibility condition again. One more differentiation with respect to time and use of the 
third equation in \rr{NS2} leads to the desired
sixth order equation
\be \label{PC3D}
\left[\nabla_2^2 + \left(1 - \frac{(2\Omega - \mathcal{S})^2}{\omega^2}\right) \partial_z^2\right] p =0. 
\ee
When $\nu_o = \nu_4 =0$ Eq. \rr{PC3D} reduces to the standard Poincar\'e-Cartan equation \rr{PCrot}. 
Eq. \rr{PC3D} is employed in the main body of this paper to determine the admissible planar wavenumber $\kappa$ values in 
equations \rr{detrot} and \rr{kappa2}.  

Eq. \rr{PC3D} can be classified according to the scheme employed by \citet[Ch. III, \S 2.3-2.6]{Courant1962} by isolating 
its principal part $\partial_z^2 \mathcal{S}^2$. Thus, the characteristic form is
\be \label{PCchar}
Q(\bm{\phi}) = \phi_3^2 \mathcal{S}^2(\bm{\phi}),
\ee
where 
\be
\mathcal{S}(\bm{\phi}) = (\nu_o-\nu_4)(\phi_1^2 + \phi^2_2) + \nu_4\phi_3^2,
\ee
$\bm{\phi} = (\phi_1, \phi_2, \phi_3)$ and the index $i=1,2,3$ denotes the variables 
$x_1=x, x_2= y, x_3 = z$ with respect to which the function $\phi$ is differentiated.
\begin{figure}
    \begin{subfigure}[t]{0.45\textwidth}
        \includegraphics[width=\linewidth]{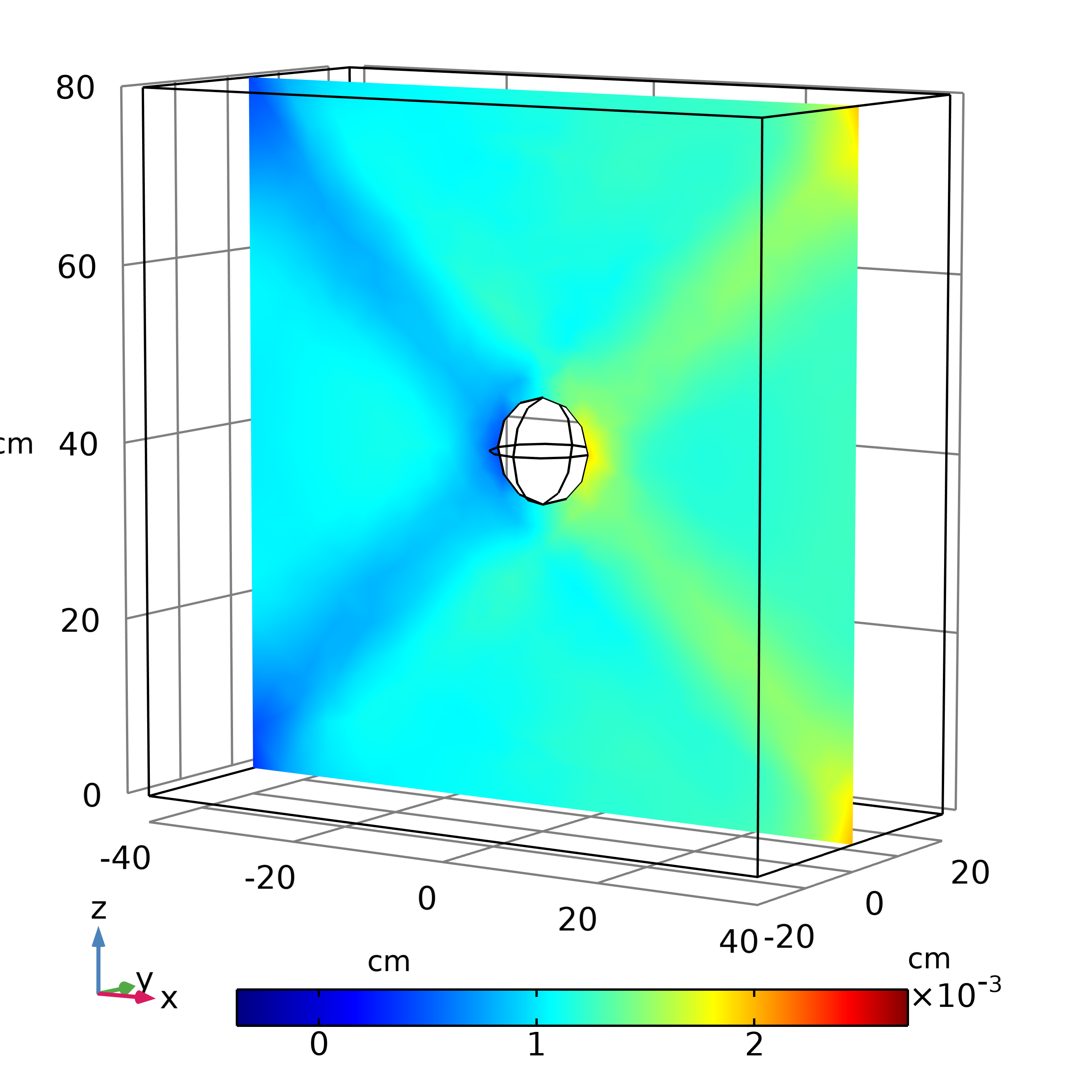} 
       \caption{ } \label{xa}
    \end{subfigure}
    \begin{subfigure}[t]{0.45\textwidth}
        \includegraphics[width=\linewidth]{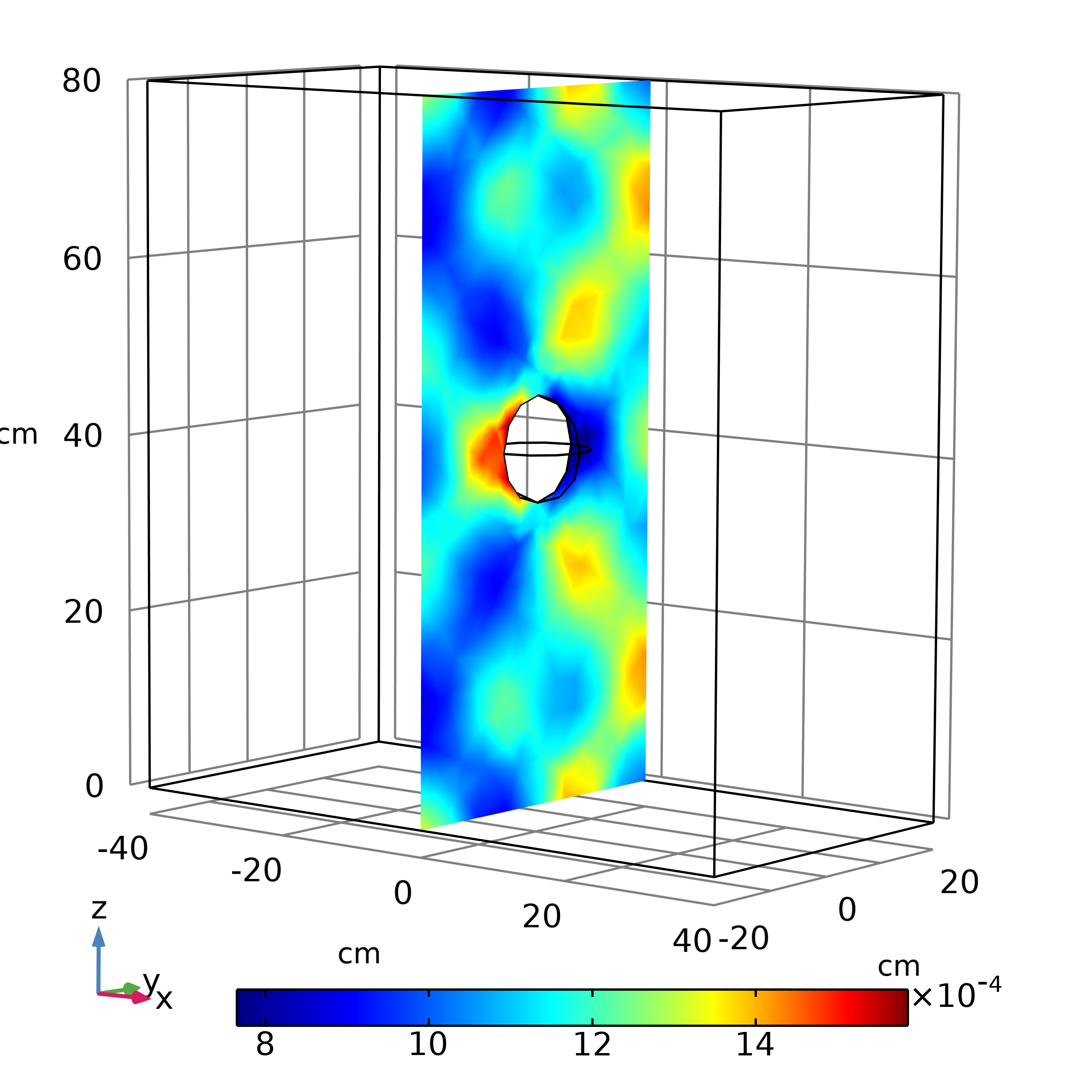} 
        \caption{}\label{xb}
    \end{subfigure}
    \caption{\label{Xpressure}Distribution of pressure (colorbar: dyne/$\textrm{cm}^2$) in an odd viscous
    liquid entering into a rectangular channel from the right, moving slowly with velocity $U=0.01$ cm/sec, and meeting a centered solid immobile sphere (of radius $6$ cm). Stokes flow with $\eta_4 = 0.2$ g/(cm sec) and $\eta_o = 0$ from the constitutive law \rr{sigma}, shear viscosity that of water and with no-slip boundary conditions on the channel walls. 
    In both panels data propagate along directions making a $45$ degree angle with the horizontal
    (along the Monge cone $z = \pm r$ in \rr{raycone}). Since the depth of the box is narrow, the data in panel (b) are reflected on its 
    walls located at $y=\pm 20$ cm. Numerical simulations were performed with the finite-element package Comsol.  
} 
\end{figure}

The characteristic form \rr{PCchar} is identical to the one obtained from Maxwell's equations \cite[p. 178]{Courant1962}
if we identify $\phi_3$ with $\tau$
(in Maxwell's case it is exact - there are no lower order terms). 
In our case there are multiple sheets $Q_1, Q_2, ...$ etc., and the characteristic form $Q$ can be expressed in the form $Q = Q_1Q_2...$, where some of the factors may be 
identical. Thus, as in \cite[p. 596]{Courant1962} rays should be defined not with reference to $Q$ but with respect to the irreducible factors $Q_j$ of $Q$. Let 
\be \label{Q1Q2}
Q_1 = \phi_3, \quad \textrm{and} \quad Q_2 = S(\bm{\phi})
\ee
be these irreducible factors. 

Consider the characteristic surface $\phi(x,y,z) = c$, where $c$ is a constant. The characteristic rays or bicharacteristics are given by 
\be \label{xidot}
\dot{x}_i = \frac{\partial Q}{\partial \phi_i} , \quad i=1,2,3,
\ee
where a superposed dot denotes differentiation with respect to some suitable curve parameter $s$.
%

For the first sheet $Q= Q_1 = \phi_3$, Eq. \rr{xidot} becomes $dx/ds=dy/ds=0$ and $dz/ds = 1$. Thus, the characteristic curves are 
\be \label{Cz}
\mathbf{C}(s) = (c_1, c_2, c_3) + (0,0,1) s, 
\ee
that is, straight lines in the $z$ direction away from a fixed point with coordinates $c_1,c_2,c_3$ and 
where $s$ is the parameter of the curve ranging along some suitable interval. 

Characteristics are also generated by $Q_2$ as defined in Eq. \rr{Q1Q2}. 
Consider first the special case when $\nu_4$ vanishes. Then, since 
$Q_2 = \phi_1^2 + \phi_2^2$, we find that characteristics are described by $\mathbf{C}(s) = (c_1, c_2, c_3) $
since $Q_2=0$ is satisfied only if $\phi_1=\phi_2=0$. 
Thus, when $\nu_4=0$, the only characteristics associated with the form \rr{PCchar} are those along the $z$ direction described in Eq. \rr{Cz}. This then
explains the presence of Taylor columns in a three-dimensional odd viscous liquid which only extend
along the $z$-direction \citep{kirkinis2023axial}.

\begin{figure}
    \begin{subfigure}[t]{0.45\textwidth}
        \includegraphics[width=\linewidth]{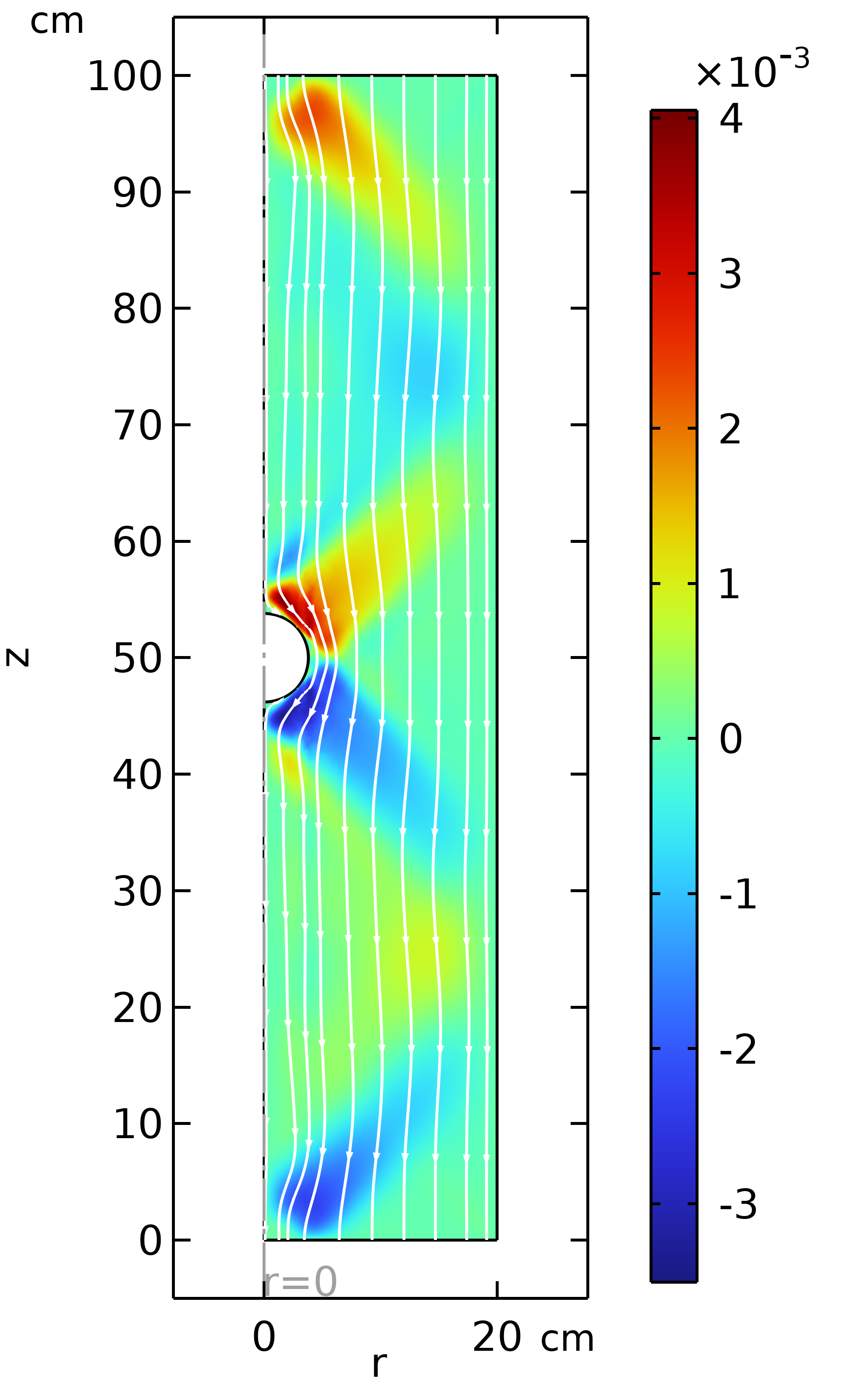} 
       \caption{Color-bar: radial velocity $v_r$ (cm/sec)} \label{vr} 
    \end{subfigure}
    \begin{subfigure}[t]{0.45\textwidth}
        \includegraphics[width=\linewidth]{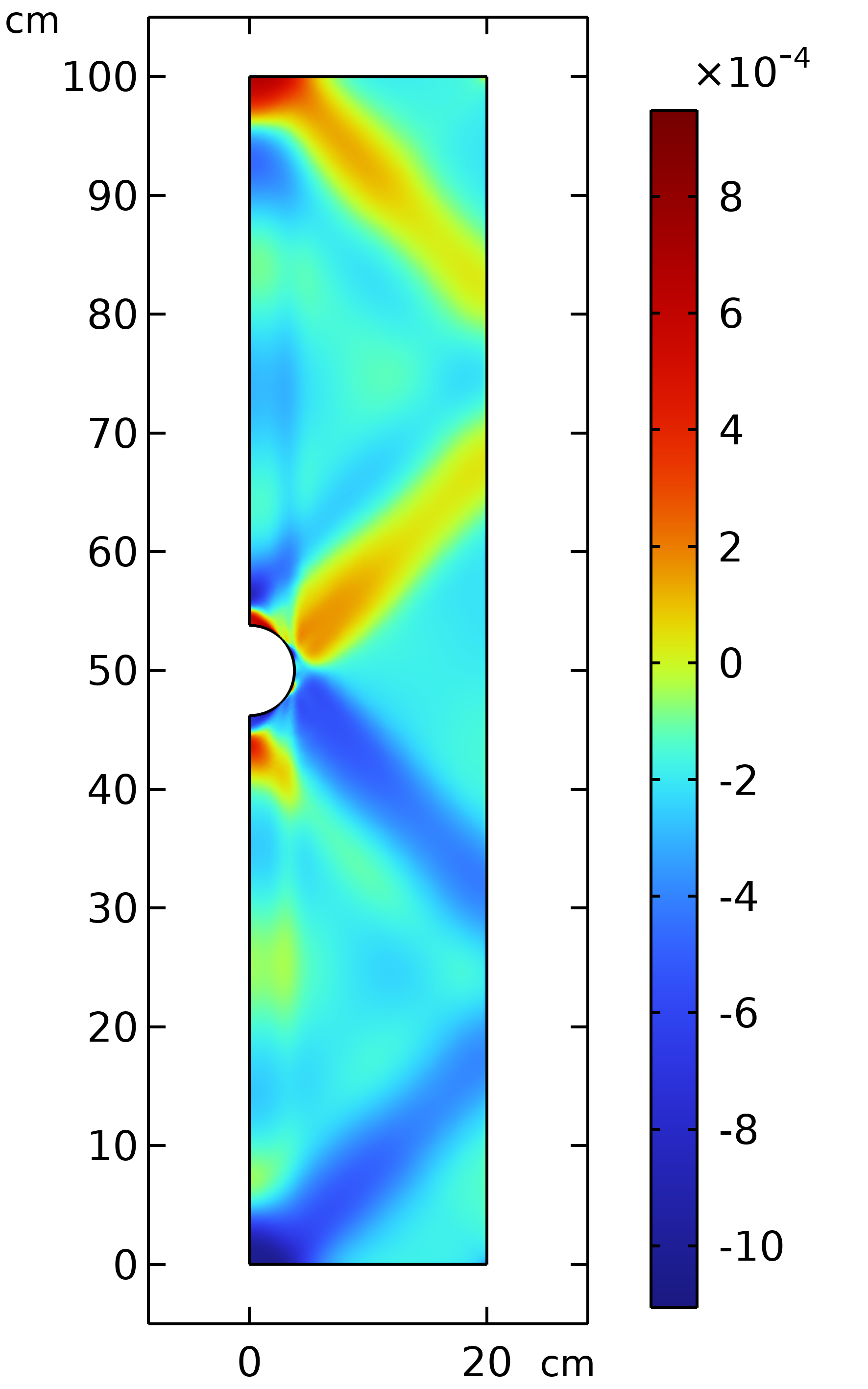} 
        \caption{Colorbar: pressure (dyne/$\textrm{cm}^2$) }\label{p}
    \end{subfigure}
    \caption{\label{cylinderp}(a) Colorbar: distribution of radial component of velocity $v_r$ (cm/sec), (where the liquid velocity is denoted by $\mathbf{v} = v_r \hat{\mathbf{r}} + 
v_\phi \hat{\bm{\phi}} + v_z \hat{\mathbf{z}}$ in cylindrical coordinates) and (b) pressure $p$ (dyne/$\textrm{cm}^2$) in an odd viscous
    liquid moving slowly and meeting an immobile sphere (of radius $3.8$ cm) located at elevation $z=50$ cm 
 at the center axis of a cylinder. Here $\eta_4 = 2$ g/(cm sec), $\eta_o = 0.1$ g/(cm sec) from constitutive law \rr{sigma}, and shear viscosity is that of water.
   Liquid enters from the top ($z=100$ cm) and exits at the bottom ($z=0$) of the cylinder. The sphere is not allowed to rotate. White lines are liquid streamlines. In all cases data emanating from the sphere propagate along rays that lie on the Monge cone $z \sim \pm r$ defined in \rr{raycone}. A column circumscribing the sphere, whose
   generators are parallel to the $z$-axis is also present. It becomes visible in a plot of the flow structure along the full expanse of the cylinder, see Fig. \ref{cylinderw}. Numerical simulations were performed with the finite-element package Comsol.  
} 
\vspace{-10pt}
\end{figure}

In the general case where both $\nu_o$ and $\nu_4$ are nonzero, $Q_2 = (\nu_o-\nu_4)(\phi_1^2 + \phi^2_2) + \nu_4\phi_3^2$ from \rr{Q1Q2} is substituted into \rr{xidot} to give
$dx/ds = (\nu_o-\nu_4)\phi_1$, $dy/ds = (\nu_o-\nu_4)\phi_2$, $dz/ds = \nu_4\phi_3$ (in each case an unimportant factor of $2$ has been absorbed into, say, $s$). Take $z$ to be the 
time-like variable and form
$\frac{dx}{dz} = \frac{ (\nu_o-\nu_4)}{\nu_4}\frac{\phi_1}{\phi_3}$ and  
$\frac{dy}{dz} = \frac{ (\nu_o-\nu_4)}{\nu_4}\frac{\phi_2}{\phi_3}$. From the condition
$Q_2=0$ we obtain $\phi_3^2 = \frac{ (\nu_4-\nu_o)}{\nu_4}(\phi_1^2 + \phi^2_2)$. 
This is only possible in the hyperbolic case, requiring that $\nu_o<\nu_4$ when $\nu_4>0$. Squaring and adding we obtain
\be
\left(\frac{dx}{dz}\right)^2  + \left(\frac{dy}{dz}\right)^2 = \frac{ (\nu_4-\nu_o)}{\nu_4}.
\ee
A solution is $(x-x_0)^2 + (y-y_0)^2 = \frac{ (\nu_4-\nu_o)}{\nu_4} z^2$ 
where $x = \alpha_1 z + x_0$ and $y=\alpha_2 z+y_0$ so, $\alpha_1^2 + \alpha_2^2 = \frac{ (\nu_4-\nu_o)}{\nu_4}$.
Thus, characteristic curves lie 
on the local ray cone or Monge cone
\be \label{raycone}
z = \pm \sqrt{\frac{\nu_4}{ \nu_4-\nu_o}} r
\ee
where $r = \sqrt{(x-x_0)^2 + (y-y_0)^2}$, 
see \cite[p. 601]{Courant1962}. 
In Fig. \ref{Xpressure} we display two slices of the domain into which liquid enters from the right wall at 
$x=40$ cm. Colorbar denotes the distribution of pressure. In both panels we've set $\nu_o =0$ and 
both panels display the direction of data along characteristics on the local ray cone \rr{raycone}
$z = \pm r$. Figure \ref{xb} shows that characteristics are reflected on the lateral channel walls located at $y=\pm 20$ cm. 
\begin{figure}
\vspace{-5pt}
\begin{center}
\includegraphics[height=3.8in,width=4in]{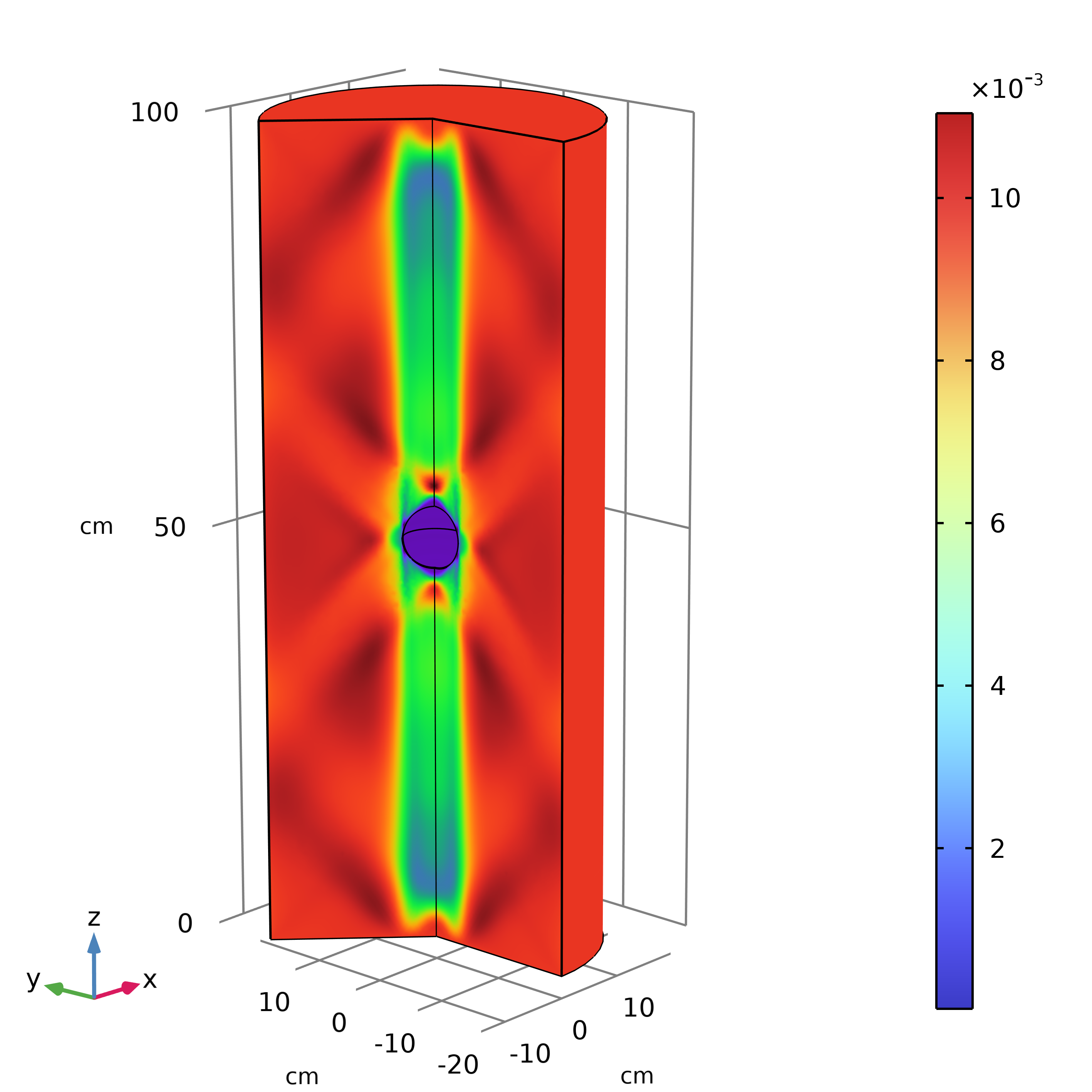}
\vspace{-0pt}
\end{center}
\caption{Colorbar: distribution of minus the axial velocity $v_z$ (cm/sec) (the z-component of the liquid velocity $\mathbf{v} = v_r \hat{\mathbf{r}} + 
v_\phi \hat{\bm{\phi}} + v_z \hat{\mathbf{z}}$ whose strength is displayed in the colorbar) in an odd viscous
    liquid moving slowly and meeting an immobile sphere (of radius $3.8$ cm) located at elevation $z=50$ cm 
 at the center axis of a cylinder (Fig. \ref{cylinderp} shows an $r$-$z$ slice of this cylinder). 
   Liquid enters from the top ($z=100$ cm) and exits at the bottom ($z=0$). The sphere is not allowed to rotate. The presence of a central Taylor column circumscribing the sphere is visible and it is attributed to the straight-line characteristics \rr{Cz} parallel to the anisotropy $z$-axis. The presence of rays making (nearly)
a 45 degree angle with the horizontal is also visible. They are attributed to the characteristics that lie
on the Monge cone \rr{raycone}. 
$\eta_4 = 2$ g/(cm sec), $\eta_o = 0.1$ g/(cm sec) from constitutive law \rr{sigma}, and shear viscosity is that of water, as in Fig. \ref{cylinderp}. Numerical simulations were performed with the finite-element package Comsol.  
\label{cylinderw}  }
\vspace{-0pt}
\end{figure}

In panel (a) of Fig. \ref{cylinderp} we display the radial component of the liquid velocity 
$\mathbf{v} = v_r \hat{\mathbf{r}} + 
v_\phi \hat{\bm{\phi}} + v_z \hat{\mathbf{z}}$
along an $r$-$z$ slice 
of a cylinder (see Fig. \ref{cylinderw} for a view of the whole cylinder) filled with odd viscous liquid with $\eta_4 = 2$ and $\eta_o=0.1$ g / (cm sec), and flowing around an immobile sphere (of radius $3.8$ cm) located at elevation $z=50$. Liquid enters from the top ($z=100$ cm) and exits at the bottom ($z=0$), with the cylinder speed (the sphere is not allowed to rotate). The white lines are the streamlines of the flow. 
We thus observe oblique propagation of data from the sphere to the wall and then their reflection. 
In panel (b) of Fig. \ref{cylinderp} we display the pressure distribution showing the same pattern of
oblique propagation of data. Since there are also characteristics that propagate vertically, 
associated with the form $Q_1$ in \rr{Q1Q2}, a Taylor-type column also exists, circumscribing 
the sphere and surrounding the central axis. This becomes visible by displaying the distribution of the axial component
of velocity $v_z$ along the whole cylinder, not just a slice. Thus, in Fig. \ref{cylinderw} we see a Taylor column
parallel to the anisotropy ($z$-axis) circumscribing the sphere, and also oblique characteristics that emanate from the sphere and propagate in a direction making an angle of (nearly) 45 degrees with the horizontal. 
Parameters employed to produce this figure:
Odd viscosity coefficent shear viscosity $\eta = 0.01$ g/(cm sec), 
cylinder radius $20$ cm, sphere radius $3.8$ cm, cylinder height $H=100$ cm, liquid density
$\rho = 1 \textrm{g/cm}^3$, liquid velocity in the $-\hat{\mathbf{z}}$ direction $U=0.01$ cm/sec. 

Summarizing, the characteristic structures that exist according to the characteristic form \rr{PCchar} are 
(assume $\nu_4>0$ for simplicity)
\begin{itemize}
\item When $\nu_o<\nu_4$ then, the operator $\mathcal{S}$ in \rr{S} is hyperbolic and characteristics exist
both in the $z$ direction (because of $Q_1$ in \rr{Q1Q2}) and the oblique direction determined by the ratio of the odd viscosities
as in Eq. \rr{raycone} .
\item When $\nu_0>\nu_4$, the operator $\mathcal{S}$ is elliptic and characteristics exist only in the 
$z$ direction (because of $Q_1$ in \rr{Q1Q2}).
\item When $\nu_0 = \nu_4$, $\mathcal{S}$ is parabolic and characteristics exist only in the 
$z$ direction (because of $Q_1$ in \rr{Q1Q2}).
\end{itemize}

We note that similar conclusions to the above were reached by employing a simpler exposure based on a generalized Taylor-Proudman
theorem \citep{Kirkinis2023halos}.

\begin{figure}
\vspace{-5pt}
\begin{center}
\includegraphics[height=2in,width=2.8in]{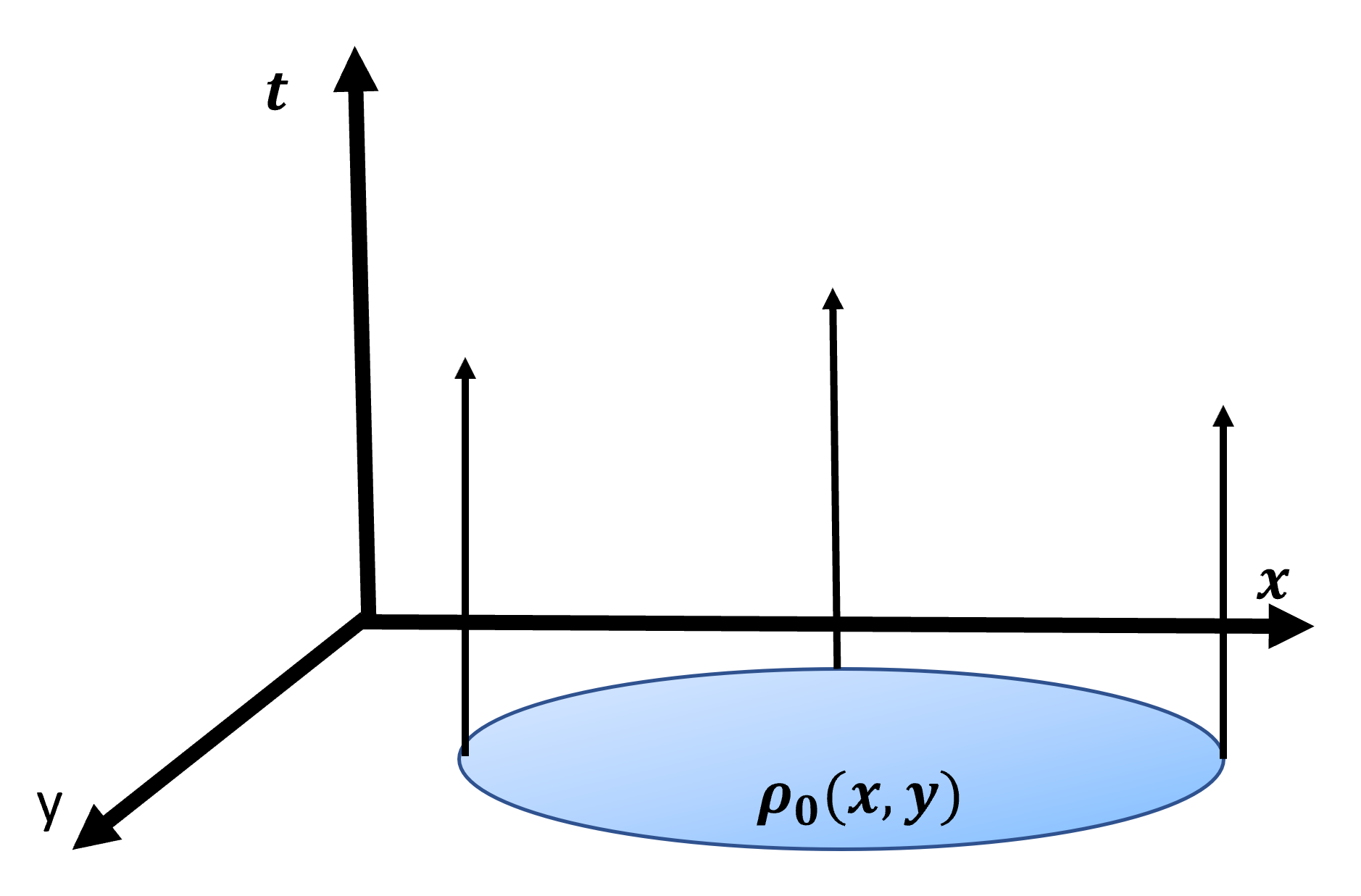}
\vspace{-0pt}
\end{center}
\caption{Initial data of density $\rho_0(x,y)$ in a compressible two-dimensional odd viscous liquid propagate along straight characteristic lines according to Eq. \rr{Q2d},
forming a temporal Taylor column. 
\label{char1}  }
\vspace{-0pt}
\end{figure}

\section{\label{sec: Poincarec}Poincar\'e-Cartan equation of a two-dimensional \emph{compressible} odd viscous liquid}
We are not aware of a Poincar\'e-Cartan equation in two-dimensional compressible flow, so we include below
the steps leading to its derivation. This equation is then employed in the main body of this paper to determine
the planar wavenumber $\kappa$ values in Eq. \rr{disp2} and \rr{kappa2c}.

Starting from the Navier-Stokes (by restoring the pressure $p = \rho' c^2$) we obtain the set of equations
\be \label{ns2d}
\partial_t u = - \frac{1}{\rho} \partial_x p + \mathcal{L}v, \quad 
\partial_t v = - \frac{1}{\rho} \partial_y p - \mathcal{L}u, \quad
\partial_t p + \rho c^2 (\partial_x u + \partial_y v) =0,
\ee
where we introduced the linear differential operator $\mathcal{L}= 2\Omega - \nu_o\nabla^2$, 
$\rho$ is a constant background density, $\rho'$ its variable part and $c$ is the speed of sound. 
We thus obtain the evolution equations
\be \label{vortdiv2d}
\partial_t(\partial_xv - \partial_y u) = - \mathcal{L}(\partial_x u + \partial_y v), \quad 
\partial_t(\partial_x u + \partial_y v) = -\frac{1}{\rho} \nabla^2 p + \mathcal{L} (\partial_xv - \partial_y u), 
\ee
for the vorticity and divergence of the velocity field. Differentiating the pressure equation in \rr{ns2d}
twice and using Eq. \rr{vortdiv2d} we obtain
\be \label{PC2Da}
\partial_t\left[ \mathcal{L}^2 - c^2 \nabla^2 + \partial_t^2   \right] p =0.
\ee
\rr{PC2Da} can be considered as the Poincar\'e-Cartan equation for a two-dimensional compressible and rigidly-rotating
odd viscous liquid. It is clear that when the pressure oscillates as $e^{-i\omega t}$ one obtains
$
\omega\left[ \mathcal{L}^2 - c^2 \nabla^2 - \omega^2   \right] =0,
$
which recovers the dispersion relation \rr{kappa2c}. 

The classification of the two-dimensional Poincar\'e-Cartan equation \rr{PC2Da} can proceed as in 
the foregoing three-dimensional case  
(see also \cite[Ch. III, \S 2.3-2.6]{Courant1962}). We replace the operators $(\partial_t, \partial_x, \partial_y)$
by time and space-like quantities $(\tau, \phi_1, \phi_2)$ and form the principal part 
\be \label{Q2d}
Q = Q_1Q_2, \quad \textrm{where} \quad Q_1 = \tau \quad \textrm{and} \quad Q_2 = \nu_0(\phi_1^2 + \phi_2^2)^2.
\ee
The situation is the same as in Eq. \rr{Cz} with the $z$ coordinate replaced by time. Initial data
propagate along vertical characteristics as displayed in Fig. \ref{char1}. This situation can also 
be understood as a \emph{temporal} Taylor column.

\bibliographystyle{jfm}
\bibliography{/Users/jmason/Desktop/LEFTERIS/Bibliography/fluids,/Users/jmason/Desktop/LEFTERIS/Bibliography/disorder,/Users/jmason/Desktop/LEFTERIS/Bibliography/materialscience,/Users/jmason/Desktop/LEFTERIS/Bibliography/perturbations,/Users/jmason/Desktop/LEFTERIS/Bibliography/Compressible,/Users/jmason/Desktop/LEFTERIS/Bibliography/physiology}

\end{document}